\documentclass[twoside,leqno]{article}
\usepackage[letterpaper]{geometry}
\usepackage{ltexpprt}
\usepackage{hyperref}

\usepackage{waingarten} 
\usepackage{tikz}
\usepackage{verbatim}
\usepackage{cancel}
\usepackage{titling}
\usepackage{microtype}
\usepackage{mleftright}
\usepackage{amsfonts}

\renewcommand{\ra}{\rightarrow}

\begin{document}


\title{\Large Nearly Tight Bounds on Testing of Metric Properties}
\author{Yiqiao Bao\thanks{University of Pennsylvania.}
\and Sampath Kannan\thanks{Simons Institute, UC Berkeley (on leave from the University of Pennsylvania).}
\and Erik Waingarten\thanks{University of Pennsylvania.}
}

\date{}

\maketitle

\begin{abstract} \small\baselineskip=9pt 
Given a non-negative $n \times n$ matrix viewed as a set of distances between $n$ points,
we consider the property testing problem of deciding if it is a metric. We also consider
the same problem for two special classes of metrics — tree metrics and ultrametrics. For general metrics, our paper is the first to
consider these questions. We prove an upper bound of $O(n^{2/3}/\eps^{4/3})$ on the query complexity for this problem.  Our algorithm
is simple, but the analysis requires great care in bounding the variance on the number of violating triangles in a sample. When
$\epsilon$ is a slowly decreasing function of $n$ (rather than a constant, as is standard), we prove a lower bound of matching dependence on $n$ of $\Omega (n^{2/3})$, ruling out any property testers with $o(n^{2/3})$ query complexity unless their dependence on $1/\eps$ is super-polynomial.

Next, we turn to tree metrics and ultrametrics. While there were known upper and lower bounds, we considerably improve these bounds showing essentially
tight bounds of $\tilde{O}(1/\eps )$ on the sample complexity. We also show a lower bound of $\Omega ( 1/\eps^{4/3} )$ on the query complexity. Our upper bounds are derived by doing a more careful analysis
of a natural, simple algorithm. For the lower bounds, we construct distributions on NO instances, where it is hard to find a witness showing that these are not ultrametrics.

\end{abstract}

\newpage
\section{Introduction}
\label{sec:intro}

Finite metric spaces is a rich topic at the intersection of combinatorics, algorithms, and geometry~(see~\cite{L02, M02, IMS17}, among many other works, for general overviews). In addition to their intrinsic interest, a strong motivation for studying metric spaces from the theoretical computer science perspective is that a metric space, or metric for short, defines a quantitative measure of dissimilarity. Once a metric over the objects, or points, is defined, one can design algorithms to find the most similar objects, partition the objects, and cluster the objects. Formally, a metric space on the ground set $[n]$ is specified by a dissimilarity function $d \colon [n] \times [n] \to \R_{\geq 0}$ which satisfies three axioms (i) the function is symmetric, so $d(i,j) = d(j, i)$, (ii) it is non-negative with $d(i,j) = 0$ if and only if $i = j$, and (iii) it satisfies the triangle inequality:
A function $d \colon [n] \times [n] \to \R_{\geq 0}$ satisfies the triangle inequality if and only if $\forall \, i, j, k \in [n]$, $d(i,k) \leq d(i,j) + d(j, k)$. 

 \emph{Property testing}~\cite{BLR93, RS96} is a natural algorithmic framework for studying whether the given matrix constitutes a metric. Here, we consider randomized $\eps$-testing algorithms that receive black-box query access to an unknown and arbitrary $n \times n$ matrix $M$. The goal is to make as few queries as possible to the entries of the matrix while deciding whether its entries encode a function $d$ which is a metric or is $\eps$-far from a metric (meaning that any metric $d' \colon [n] \times [n]\to \R_{\geq 0}$ differs on at least $\eps$-fraction of the $n^2$ inputs). We remark that ``$\eps$-farness'' here, which is the most natural from a property testing perspective, provides an $\ell_0$-distance guarantee, which has seen a recent surge of interest in approximation algorithms~\cite{FRB18, CFLM22, CG23, K23}. Our work gives upper and lower bounds on various metric testing problems and
 can be used as a \emph{sublinear}-time preprocessing step, or quick ``sanity check,'' to the more expensive polynomial-time approximation algorithms. 

As we detail in Subsection~\ref{sec:related-work}, there are a number of prior works that study finite metrics from a property testing perspective. Despite the broad interest, this work is the first to address what is arguably the most basic metric testing question---can finite metrics be efficiently tested? Our main algorithmic result is a non-adaptive metric testing algorithm that makes $O(n^{2/3}/\eps^{4/3})$ queries. As is common in property testing, the algorithm is straightforward and proceeds by executing two checks: 
\begin{enumerate}
    \item It randomly selects a subset of $O(1/\eps)$ points from $[n]$ and a subset of $O(n^{2/3}/\eps^{1/3})$ pairs from $[n]\times [n]$, and checks whether the triangle formed by a point and pair violates the triangle inequality.
    \item It samples a random set of $O(n^{1/3}/\eps^{2/3})$ many points and checks whether there are three points in this set that violate the triangle inequality.
\end{enumerate}
Every metric space will trivially pass the above tests since there are no violations of the triangle inequality. The interesting aspect of our analysis, which constitutes the majority of the technical work, establishes that if no violations of the triangle inequality are observed with probability $1/3$, the function $d \colon [n]\times [n] \to \R_{\geq 0}$ is (almost) a metric. Furthermore, we show that a lower bound establishing that (unless the dependence on $\eps$ becomes super-polynomial) a dependence on $n^{2/3}$ is needed for non-adaptive algorithms with one-sided error.

Next, we turn our attention to the specific classes of tree metrics and ultrametrics. These were first studied by Parnas and Ron~\cite{PR01} from the property testing perspective and also received recent attention from the above-mentioned works on approximation algorithms. Tree metrics (also known as additive metrics) and ultrametrics are specific sub-classes of metric spaces that may be represented with positively-weighted trees. Points in the metric correspond to nodes in a tree and distances are measured by the lengths of paths (see Section~\ref{sec:prelims} for their formal definition). In their work,~\cite{PR01} give an algorithm which samples $O(1/\eps^3)$ random points and queries all pairwise function evaluations (using $O(1/\eps^{6})$ queries) to $\eps$-test tree metrics and ultrametrics. We improve upon their dependence in the following way: we show that, for both tree metrics and ultrametrics, it suffices to take $\tilde{O}(1/\eps)$ samples and query all $\tilde{O}(1/\eps^2)$ pairwise evaluations. Furthermore, we show a sample complexity lower bound of $\Omega(1/\eps)$ (i.e., testing algorithms must interact with at least these many points), and a $\Omega(1/\eps^{4/3})$ lower bound on the query complexity. Hence, our algorithms are sample-optimal (up to $\polylog(1/\eps)$ factors), and the $\Omega(1/\eps^{4/3})$ query lower bound rules out what is often the ``best case'' scenario in property testing, which is an $\Theta(1/\eps)$-query tester. 

\subsection{Our Contributions}

\paragraph{Testing Metrics.} As mentioned above, our work is the first to prove upper and lower bounds on the query complexity of property testing of metric spaces. We provide three (non-adaptive) testing algorithms for testing metrics, tree metrics, and ultrametrics, along with lower bounds which show that our results are (in certain regimes) the best possible. We begin by stating our main algorithmic theorem, which gives an algorithm for testing metrics using $O(n^{2/3} / \eps^{4/3})$ queries.

\begin{theorem}[Testing Metrics---Upper Bound]\label{thm:testing-metric}
For any large enough $n \in \N$ and any $\eps \in (0,1)$, there exists a randomized algorithm that receives query access to an unknown matrix $M \in \R^{n\times n}$ and makes $O(n^{2/3} / \eps^{4/3})$ queries with the following guarantee:
\begin{itemize}
\item If $M$ defines a metric space on $[n]$, the algorithm outputs ``accept'' with probability $1$.
\item If $M$ is $\eps$-far from being being a metric, then the algorithm outputs ``reject'' with probability at least $2/3$.
\end{itemize}
Furthermore, the algorithm is non-adaptive (i.e., queries made do not depend on answers to prior queries).
\end{theorem}

The algorithm that proves Theorem~\ref{thm:testing-metric} is especially appealing from a structural perspective due to its simplicity. By executing only two types of checks and only $O(n^{2/3}/\eps^{4/3})$ queries, the algorithm guarantees that any matrix that passes the test is $\eps$-close to a metric with high probability. Our second result shows that the dependence on the number of points must be $n^{2/3}$ unless one is willing to incur super-polynomial factors in $\eps$. 

\begin{theorem}[Testing Metrics---Lower Bound]\label{thm:testing-metric-lb}
For any large enough $n \in \N$, let $\eps = n^{-\nu(n)}$ where $\nu(n) = (\log \log \log n + 4) / \log\log n$. Any non-adaptive, one-sided algorithm which can $\eps$-test whether a matrix $M \in \R^{n\times n}$ encodes a metric must make $\Omega(n^{2/3 + 2\nu(n)/3})$ queries. 
\end{theorem}

The lower bound implies that, unless the dependence on $\eps$ blows up and becomes super-polynomial (in particular, at least $(1/\eps)^{\Omega(1/\nu(n))}$, then a dependence of $n^{2/3}$ is unavoidable. In other words, one cannot hope for an algorithm whose complexity is $O(n^{.65} / \eps^{c})$ for any fixed constant $c > 0$. The proof stems from a connection between testing metric spaces and triangle-freeness testing; we will construct matrices $M \in \R^{n\times n}$, which masquerade as metrics to low-query algorithms, by utilizing certain Behrend graphs that have previously appeared for proving lower bounds on testing triangle-freeness in graphs~\cite{A02, AKKR08}. That work, in Section~6, shows a reduction from one-sided triangle-freeness testers to two-sided triangle freeness testers, which we believe applies mutatis mutandis, so $\Omega(n^{2/3+2\nu(n)/\eps})$ queries are necessary for two-sided non-adaptive metric testing algorithms as well. 

We remark that the connection between triangle-freeness testing and metric testing is useful for lower bounds, but not for our upper bounds. As we will discuss, our proof of Theorem~\ref{thm:testing-metric} relies on certain properties of metrics that do not have a direct analog in graphs. 

\paragraph{Tree Metrics and Ultrametrics.} We now state the main results for testing tree metrics and ultrametrics. As mentioned, our works improve on the algorithms of~\cite{PR01} by improving the sample complexity from $O(1/\eps^3)$ to $\tilde{O}(1/\eps)$, and the query complexity from $O(1/\eps^6)$ to $\tilde{O}(1/\eps^2)$. Our lower bounds establish the following two aspects: (i) our algorithms are sample optimal, meaning any $\eps$-testing must evaluate distances to $\Omega(1/\eps)$ many points, and (ii) one cannot hope for the ``best case'' scenario of a $O(1/\eps)$-query tester. 

\begin{theorem}[Testing Tree Metrics and Ultrametrics---Upper bound]\label{thm:testing-trees}
For any large enough $n\in \N$ and any $\eps > 0$, there exists a randomized algorithm that receives query access to an unknown matrix $M \in \R^{n\times n}$, using $\tilde{O}(1/\eps)$ samples and $\tilde{O}(1/\eps^2)$ queries, and has the following guarantee:
\begin{itemize}
\item If $M$ defines a tree metric (or ultrametric) on $[n]$, the algorithm always outputs ``accept.'' 
\item If $M$ is $\eps$-far from being a tree metric (or ultrametric) on $[n]$, the algorithm outputs ``reject'' with probability at least $2/3$.
\end{itemize}
Furthermore, the algorithm is non-adaptive. 
\end{theorem}

Our algorithm is exactly the same as the algorithm of~\cite{PR01}, and the improvement lies solely in the analysis. The algorithm takes $\tilde{O}(1/\eps)$ samples and queries all pairwise evaluations using $\tilde{O}(1/\eps^2)$ queries. We show that, if there are no violations to tree metrics (or ultrametrics) in a sample of size $\tilde{O}(1/\eps)$ with probability at least $1/3$, then $M$ is $\eps$-close to a tree metric (or ultrametric).

\begin{theorem}[Testing Tree Metrics and Ultrametrics---Lower bound]\label{thm:trees-lb}
For any large enough $n \in \N$ and $\eps \in (0, 1)$, any non-adaptive algorithm which can $\eps$-test whether a matrix $M \in \R^{n\times n}$ is a tree metric (or ultrametric) must use $\Omega(1/\eps)$ samples and at least $\Omega(1/\eps^{4/3})$ queries. 
\end{theorem}

\subsection{Related Work}\label{sec:related-work}

There are a number of works on sublinear algorithms for metric spaces. The most relevant to this work is~\cite{PR01} who, among other results, gave property testing algorithms for tree metrics and ultrametrics using $O(1/\eps^3)$ samples and $O(1/\eps^6)$ queries. More generally viewing metrics as a property of $n\times n$ matrices, there has been a recent line-of-work on testing of matrix properties~\cite{FN01, BLWZ19,BCJ20, B23,BDDMR24} (see, also Chapter 8 in the textbook~\cite{BY22}). The works of~\cite{KS03, O08} study property testing of points in a metric, where the algorithm receives query access to a distance oracle, and the proximity parameter $\eps$ is with respect to the number of points which must be modified; the properties of interest are dimensionality and embeddability into other metrics. Property testing by accessing vectors directly (for example, when the metric is Euclidean) has also been studied~\cite{CSZ00, CS01}, and more generally, there have been various sublinear algorithms in these settings~\cite{CLM03, BCIS05, CS09, CS20}. 

A recent line of work in approximation algorithms, under the name ``metric violation distance,'' finds the (approximately) most similar metric (with respect to the $\ell_0$-distance) in polynomial time~\cite{FRB18, CFLM22, CG23, K23}. There, the $\ell_0$-distance is the number of entries of $d \colon [n]\times [n] \to \R_{\geq 0}$ which are changed, coinciding with the proximity parameter $\eps$ studied in property testing. For example, the current best algorithm~\cite{CFLM22} runs in $O(n^{3})$ time and produces a metric space which is a $O(\log n)$-multiplicative factor further than the closest metric. In contrast, property testing is much more efficient but only provides approximate decisions. Hence, our property testing algorithms (both for general metrics, as well as tree metrics or ultrametrics) could be used as a preprocessing step---if a property tester declares a supposed distance-matrix is already too far from a metric (or tree or ultrametric), there may be no use in more expensive approximation algorithms.

\subsection{Technical Overview}

\paragraph{Metric Testing Upper Bound.} \ignore{We first overview the techniques involved in proving Theorem~\ref{thm:testing-metric}, where we take $\eps$ to be a small constant to highlight the $n^{2/3}$-dependence. Our goal is to execute two types of checks (Figure~\ref{fig:metric-testing}), which will sample points and determine whether there are violations of the triangle inequality among the sampled points. The algorithm will output ``reject'' only if a violation of the triangle inequality is found, so the entire analysis consists of showing that if a matrix passes the test with probability at least $1/3$, then it must be close to a metric, or equivalently, if a matrix $M$ is $\eps$-far from a metric, then a violation to the triangle inequality is found with probability at least $2/3$.\footnote{The conditions of symmetry, non-negativity, and zeros in the diagonal are easily handled (see Lemma~\ref{lem:clean}.}}

Our starting point is that, any matrix $M$ which is $\eps$-far from being a metric must contain $\Omega(\eps n^2)$ triangles of three points $\{i,j,k\}$ whose pairwise distances violate the triangle inequality (Lemma~\ref{lem:exist}). These so-called ``violating triangles,'' denoted by the collection $T$, are evidence that $M$ is not a metric, and our algorithm's goal is to query all three entries of at least one violating triangle if they exist. This already suggests an $O(n^{2/3})$-query tester (assume, for this overview, that $\eps= \Omega(1)$) since the expected number of triangles from $T$ contained within a sample $\bi_1,\dots, \bi_s \sim [n]$ is
\[ |T| \cdot \Prx\left[ \text{a fixed triangle is among $s$ random points}\right] = \Omega(\eps n^2) \cdot \Omega\left(\frac{s^3}{n^3} \right),\]
which is a large constant when the size sample $s$ is $\Theta(n^{1/3})$, and leads to an $O(n^{2/3})$ query complexity. The challenge is upper bounding the variance of this random variable. For example, if all violating triangles are incident on a single vertex or on $O(n)$ pairs of vertices, sampling these requires $\Omega(n)$ queries. Our analysis will rule out such cases by showing matrices $M$, with these triangle configurations, are close to a metric (even if there are many violating triangles). Furthermore, we show that as long as $M$ is $\eps$-far from being a metric, one of two things happens:
\begin{itemize}
\item Either there are $\Omega(\eps n)$ points, each of which participates in $\Omega(n^{4/3})$ violating triangles, or
\item There exists a structured set of $\Omega(\eps n^2)$ violating triangles $\tilde{T}$ where each point participates in at most $O(n^{4/3})$ triangles, and each pair of points in at most $O(n^{2/3})$ triangles.
\end{itemize} 
In the first case, we run our first algorithmic check: sample $O(1/\eps)$ points and check the triangle inequality against $O(n^{2/3})$ random pairs (which guarantees one of the $\Omega(\eps n)$ points and corresponding $\Omega(n^{4/3})$ pairs is sampled, see Lemma~\ref{lem:check-hi-degree}). Proving the second case is more challenging (Lemma~\ref{lem:goodSet}), but once we do, the bounds on the number of participating triangles on points and pairs of points help us prove an adequate bound on the variance of the number of violating triangles ``caught'' by a random $O(n^{1/3}$)-size sample (see Claim~\ref{cl:exp-var}).

The second case, Lemma~\ref{lem:goodSet}, proceeds in the following way. Suppose $M$ is $\eps$-far from being a metric and let $T$ be the violating triangles. If $o(\eps n)$ points participate in more than $\Omega(n^{4/3})$ violating triangles, we may ``fix'' all such points---we change all $O(n)$ entries incident to each of these points for a total of $o(\eps n^2)$ entries (Lemma~\ref{lem:vertex-deg}). After these changes, the new matrix $M'$ is still $\eps/2$-far from a metric, but each point now participates in $O(n^{4/3})$ violating triangles. We then address the pairs of points (corresponding to entries in $M'$) that participate in too many, $\Omega(n^{2/3})$, violating triangles. We show that there are $o(\eps n^2)$ such pairs since each point participates in few violating triangles (Lemma~\ref{lem:low-heavy-edge}), but we may not be able to ``fix'' these entries---note that there are $n-2$ other points whose distances constrain that of the pair, and these constraints may be irreconcilable by changing only one entry.

Here is where the fact that metrics specify real-valued distances (as opposed to graphs, which only specify Boolean values) becomes useful. Suppose a pair of points $\{i,j\}$ participates in two violating triangles, with points $k_1$ and $k_2$ respectively, and further, one cannot change the distance between $(i,j)$ and simultaneously ``fix'' the triangles $\{i,j,k_1\}$ and $\{i,j,k_2\}$. Then, we show that either $\{i, k_1, k_2\}$ or $\{j, k_1, k_2\}$ must also be a violating triangle (Claim~\ref{c:intervals}). This implies that, since $i$ and $j$ each participates in $O(n^{4/3})$ violating triangles, there must exist a setting which fixes all but $O(n^{2/3})$ violating triangles incident on $i,j$ (Lemma~\ref{lem:modify}). After performing this modification, the resulting matrix $M''$ is still $\eps/4$-far, and so must contain $\Omega(\eps n^2)$ violating triangles and both bounds on the number of violating triangles on points and pairs of points hold. These violating triangles may not exist in $M$, since they may use modified entries, but we show how to map such violating triangles back to violating triangles that do exist in $M$. (The assumption was that the modified pairs of points participated in many violating triangles and so there will be many such distinct triangles to choose from.) This completes the analysis of the metric testing algorithm.

\paragraph{Metric Testing Lower Bound.} The lower bound on metric testing comes from a connection to triangle-freeness testing in graphs. We base our lower bound on the construction in~\cite{AKKR08}, who give tripartite graphs on $3n$ vertices and $\Theta(n^{2-\nu(n)})$ edge-disjoint triangles, so-called Behrend graphs. Our task, which we do in Claim~\ref{cl:matrix-construction}, is to assign weights to edges (as well as non-edges) of the Behrend graph so that the $\Theta(n^{2-\nu(n)})$ triangles are the only violations to the triangle inequality. A non-adaptive, one-sided lower bound of $\Omega(n^{2/3 + 2\nu(n)/3})$ proceeds as follows. Fix $q$ queries to entries of a $(3n) \times (3n)$ matrix, and let $\bM$ denote the (random) Behrend-based matrix obtained after randomly permuting points. In order to observe a violation of the triangle inequality, there must be a triangle formed by the $q$ queries which contains one of the $\Theta(n^{2-\nu(n)})$ edge-disjoint triangles of the Behrend graph. Among $q$ queries over pairs, there can be at most $O(q^{3/2})$ triangles (because cliques of size $\Theta(\sqrt{q})$ maximize the number of triangles with $q$ edges), and the probability that a fixed triangle of the Behrend graph is mapped to a fixed triangle among the queries under a random permutation is $O(1/n^3)$. Hence, the probability we observe at least one triangle is at most
\[ O(q^{3/2}) \cdot O(n^{2-\nu(n)}) \cdot O(1/n^{3}) = O(q^{3/2} / n^{1+\nu(n)}), \]
which is $o(1)$ when $q$ is $o(n^{2/3 + 2\nu(n)/3})$. This establishes the one-sided lower bound; to make the lower bound two-sided, there is a reduction from~\cite{AKKR08} which applies for graphs, and readily applies in our scenario as well.

\paragraph{Tree Metric and Ultrametric Testing Upper Bound.} Testing tree metrics and ultrametrics is significantly more efficient, because tree metrics and ultrametrics are more constrained than (general) metrics, making violations easier to find. Consider ultrametrics, which satisfy the following strengthening of the triangle inequality: for any $i,j,k \in [n]$, if $d(i,j)$ is the maximum distance among the three pairs, then
\[ d(i,j) = \max\left\{ d(i,k), d(j,k) \right\}. \]
Our analysis mirrors that of~\cite{PR01} (which obtained a $O(1/\eps^6)$ query complexity) and incorporates one change to obtain the query complexity $\tilde{O}(1/\eps^2)$. Roughly speaking, a set of sampled points $\bS$ which does not contain a violation amongst themselves imposes constraints on distances amongst the remaining points in $[n] \setminus \bS$. For example, if $j$ and $k$ are two points in $[n] \setminus \bS$, and some $i \in \bS$ satisfies $M(i,j) \neq M(i,k)$, an ultrametric \emph{must set} $M(j,k)$ to $\max\{ M(i,j), M(i,k) \}$; a violation occurs when it does not. Hence, consider the partition of $[n] \setminus \bS$ imposes by $\bS$, where two points $j,k \in [n]\setminus \bS$ belong to different parts if $M(i,j) \neq M(i,k)$ for some $i \in \bS$ (Definition~\ref{def:skeleton-separator}). As parts become smaller, there are more pairs $(j,k)$ in different parts, which makes violations easier to find.~\cite{PR01} argues (using a loose argument which does not optimize $\eps$-factors) that a batch of $O(1/\eps)$ samples added to $\bS$ decreases the number of pairs in the same part by $\Omega((\eps n)^2)$, and the analysis follows since one cannot decrease a count of pairs (which are at most $n^2$ and always non-negative) by $\Omega((\eps n)^2)$ more than $O(1/\eps^2)$ times (see proof of Theorem~3 in~\cite{PR01}). We follow the same plan but show that the expected number of pairs in the same part after a single sample decreases by a multiplicative $(1 - \Omega(\eps))$-factor (Lemma~\ref{lem:active-entry-reduce_AS})---the improved bound of $O(\log(1/\eps) / \eps)$ sample complexity follows analogously.

\paragraph{Tree Metric and Ultrametric Testing Lower Bound}

We prove a sample complexity lower bound of $\Omega (1/\eps)$ for one-sided error testing algorithms by constructing a distribution over matrices that are far from ultrametrics and tree metrics, where each violating triple contains one out of a set of $\epsilon n$ points, making it necessary to sample at least one of these vertices in order to detect a violating triple.

We prove a query complexity lower bound of $\Omega (1/\eps^{4/3})$ for testing ultrametrics, which in turn implies the same query lower bound for tree metrics. This is done by constructing a different distribution where we partition $[n]$ into
$1/\eps$ groups each of size $\eps n$, such that any violation involves all 3 points from the same group. We also prove that to maximize the probability of finding a violation, a tester is best off making queries on every pair of points in a suitably chosen sample. Finally, we show that to have a decent probability of finding a violation, the sample size must be $\Omega (1/\eps ^{2/3})$, leading to the query lower bound of $\Omega (1/ \eps^{4/3})$.
\section{Preliminaries}\label{sec:prelims}

We use the standard definitions of metrics, tree metrics, and ultrametrics. In order to be self-contained, we include these definitions in the appendix.

Next, we recall the standard model of property testing which we will use throughout the paper. We state the definitions of property testing as testing properties of matrices. As we will see, a distance function $d \colon [n] \times [n] \to \R_{\geq 0}$ may be encoded by an $n\times n$ matrix. This view will be useful, as our analysis will refer to ``blocks'', and ``block-diagonal'' structures which are more intuitive for matrices. 

Recall that the goal of property testing algorithms is to provide very efficient, \emph{sublinear time} query algorithms which approximately decide whether an object satisfies a property or is $\eps$-far from satisfying a property. Towards that end, we consider algorithms that test with respect to the $\ell_0$-distance which counts the number of entries (i.e., coordinates) where two matrices differ. Throughout the paper, we will encode functions $d \colon [n] \times [n] \to \R$ as $n\times n$ matrices. 
\begin{Definition}
Given two $n \times n$ matrices, $A, B \in \R^{n\times n}$, let
\[ \| A - B \|_0 = \sum_{i=1}^n \sum_{j=1}^n \ind\{ A_{ij} \neq B_{ij} \}. \]
We will refer to a property $\calP$ of $n\times n$ matrices by letting $\calP$ denote the subset of matrices that satisfy the property. For a subset $\calP$ of $n \times n$ matrices and any $\eps > 0$, we say that an $n \times n$ matrix $M$ is $\eps$-far from $\calP$ if
\[ d_{\ell_0}(M, \calP) = \inf_{A \in \calP} \| M - A \|_0 \geq \eps n^2. \]
 
\end{Definition}

\begin{Definition}[Property Testing Algorithm for $n\times n$ Matrices]
For any $n \in \N$ and $\eps > 0$, an $\eps$-testing algorithm which tests a property $\calP$ is a randomized algorithm that receives query access to an unknown matrix $M \in \R^{n \times n}$ and satisfies:
\begin{itemize}
\item \emph{\textbf{Completeness}}: If $M \in \calP$, the algorithm outputs ``accept'' with probability at least $2/3$.
\item \emph{\textbf{Soundness}}: If $M$ is $\eps$-far from $\calP$, the algorithm outputs ``reject'' with probability at least $2/3$. 
\end{itemize}
The algorithm is non-adaptive if the queries made do not depend on the answer to prior queries (i.e., all queries are made in parallel), and the algorithm achieves one-sided error if it accepts inputs $M \in \calP$ with probability $1$. We design algorithms that measure the following notions of complexity:
\begin{itemize}
\item \emph{\textbf{Sample Complexity}}: The sample complexity of an algorithm is the maximum number of distinct rows or columns of the input matrix that are queried by the algorithm.
\item \emph{\textbf{Query Complexity}}: The query complexity of an algorithm is the maximum number of entries of the input matrix that are queried by the algorithm.
\end{itemize}
\end{Definition}

The definition of a metric requires the distance function to be non-negative with $d(i,i) = 0$, symmetric, and satisfy the triangle inequality (or the stronger four-point-condition and three-point-condition in metrics and ultrametrics). Our tests and analysis, which receive as input an arbitrary $n\times n$ matrix, will assume that the matrix is non-negative, symmetric, and zero only along the diagonal. The reason for this assumption is that these conditions may be easily checked algorithmically and incorporated into our testing algorithm. Formally, we state the following lemma, which allows us to design testing algorithms while focusing solely on the ``interesting'' violations; the lemma follows trivially by checking $O(1/\eps)$ randomly chosen entries. 
\begin{lemma}\label{lem:clean}
For $n \in \N$, let $\calP$ denote any property of $n\times n$ matrices, and let
\[ \calC \eqdef \left\{ M \in \R^{n \times n} : \begin{array}{l} \text{(i) for all $i,j \in[n]$, $M(i,j) = M(j,i)$} \\
		\text{(ii) for all $i,j \in [n]$ with $i\neq j$, $M(i,j) > 0$} \\
		\text{(iii) for all $i\in[n]$, $M(i,i) = 0$} \end{array} \right\}. \]
Suppose there is an $\eps$-testing algorithm for $\calP$ which assumes the input $M \in \calC$ and uses $q(n,\eps)$ queries. Then, there is an $\eps$-testing algorithm for $\calP \cap \calC$ using $O(q(n, \eps)) + O(1/\eps)$ queries. 
\end{lemma}


\newcommand{\CheckHiDegree}{\textsc{CheckHiDegree}}
\newcommand{\CheckViolation}{\textsc{CheckViolation}}
\newcommand{\VaSampling}{\textsc{VanillaSampling}}
\newcommand{\fT}{\tilde{T}}
\newcommand{\sco}{4}
\renewcommand{\lsim}{{\;\raise0.3ex\hbox{$<$\kern-0.75em\raise-1.1ex\hbox{$\sim$}}\;}}
\newcommand{\purple}{\color{purple}}
\newcommand{\ea}{\eps^{1/3}}
\newcommand{\eb}{\eps^{-1/3}}

\section{Metric Testing Upper Bound: Theorem~\ref{thm:testing-metric}} \label{sec:general}

In this section, we present an algorithm for testing general metrics with query complexity 
$O(n^{2/3}/\epsilon^{4/3})$, thereby proving Theorem~\ref{thm:testing-metric}. Throughout this section, we will consider $n \in \N$ as the main asymptotic parameter, and let
\[ \calP = \left\{ M \in \calC : M \text{ encodes a metric space over $[n]$}\right\}. \]
Recall that the set $\calC$ consists of $n\times n$ matrices which we call \emph{clean}, meaning that they are symmetric, non-negative, and zero if and only if on the diagonal. The algorithm is straightforward and executes two types of checks, attempting to find violations of the triangle inequality. We begin by stating the definition which will be needed for the analysis.

\begin{Definition}
A triangle is a set $\{i,j,k\}$ of three distinct indices $i, j, k \in [n]$. A triangle $\{i,j,k\}$ is \emph{violating} for $M \in \calC$ if it forms a violation of the triangle inequality. Namely, after re-naming indices so $M(i,j)$ is the maximum among pairwise evaluations,
\[ M(i,j) > M(i, k) + M(k, j). \]
\end{Definition}

\begin{lemma}\label{lem:exist}
For any $\eps \in (0, 1)$, and any $M\in \calC$ which is $\eps$-far from $\calP$, there are at least $\eps n^2/6$ distinct violating triangles $\{i,j,k\}$ of $M$.
\end{lemma}
\begin{proof} See appendix. \end{proof}

\begin{Definition}[Triangle Degree]\label{def:hypergraph}
For $M \in \calC$ and a set of violating triangles $T$ of $M$, we let $B(T) = ([n], T)$ be the $3$-uniform hypergraph whose hyper-edges are violating triangles. 
\begin{itemize}
\item For an index $i \in [n]$, the \emph{vertex-triangle-degree}, $d_{T}(i)$ is the number of violating triangles containing $i$,
\[ d_{T}(i) = \left| t \in T : i \in t \right|. \]
\item For a pair $(i,j) \in [n]\times [n]$, the \emph{edge-triangle-degree}, $d_{T}(i,j)$ is the number of violating triangles containing both $i$ and $j$,
\[ d_{T}(i,j) = \left| t\in T : i,j \in t \right|. \]
\end{itemize}
\end{Definition}

Our test proceeds by executing two sub-routines $\CheckHiDegree$ and $\CheckViolation$ which aim to find a violating triangle. We then state the two lemmas concerning these sub-routines and show how they imply the main tester.

\begin{framed}
    \textbf{Metric Testing Algorithm.} The algorithm will aim to find a triangle $\{i,j,k\}$ which forms a violation of the triangle inequality. If it does find one, a violation is a certificate that the input matrix is not a metric and the algorithm will output ``reject.''
    
    \textbf{Input}: The parameters $n \in \N$ and $\eps \in (1/n,1)$, as well as query access to the entries of an unknown $n \times n$ matrix $M$ from $\calC$ (see Lemma~\ref{lem:clean}).  For $\eps < 1/n$, the claimed complexity $O(n^{2/3} /\eps^{4/3})$ becomes $O(n^2)$, so one can read the entire matrix.\\ 
    \textbf{Output}: ``accept'' or ``reject.''
    
    \begin{enumerate}
    \item Execute \CheckHiDegree$(M, \eps)$. If the sub-routine outputs ``reject,'' then output ``reject.''
    \item Execute \CheckViolation$(M, \eps)$. If the sub-routine outputs ``reject,'' then output ``reject.''
    \item If neither sub-routine has output ``reject,'' then output ``accept.''
    \end{enumerate}
\end{framed}


\begin{framed}
    \textbf{The $\CheckHiDegree$ Sub-routine.} 
    
    \textbf{Input}: The parameters $n \in \N$ and $\eps \in (1/n,1)$, as well as query access to the entries of an unknown $n \times n$ matrix $M \in \calC$. \\ 
    \textbf{Output}: ``accept'' or ``reject.''
    \begin{enumerate}
    \item For $u = O(1/\eps)$, take $u$ random samples $\bi_1,\dots, \bi_u \sim [n]$ drawn independently. For $s=O(n^{2/3}/\eps^{1/3})$, take $s$ random sample pairs $(\bj_1,\bk_1)\dots, (\bj_s,\bk_s) \sim [n]^2$ drawn independently. Query $M(\bi_{\ell}, \bj_{t}),M(\bi_{\ell}, \bk_{t}),M(\bj_{t},\bk_{t})$ for all $\ell\in[u], t \in [s]$. 
    \item If there exists a triangle among the sampled indices $\{ \bi_{\ell}, \bj_{t}, \bk_{t} \}$ which is a violating triangle in $M$, output ``reject.'' Otherwise, output ``accept.''
    \end{enumerate}
    
    \textbf{The $\CheckViolation$ Sub-routine.}
    
    \textbf{Input}: The parameters $n \in \N$ and $\eps \in (1/n,1)$, as well as query access to the entries of an unknown $n \times n$ matrix $M \in \calC$. \\ 
    \textbf{Output}: ``accept'' or ``reject.''

    \begin{enumerate}
    \item For $s = O(n^{1/3} / {\eps^{2/3}})$, take $s$ random samples $\bi_1,\dots, \bi_s \sim [n]$ drawn independently. Query $M(\bi_{\ell}, \bi_{k})$ for all $\ell, k \in [s]$. 
    \item If there exists a triangle among the sampled indices $\{ \bi_{\ell}, \bi_{k}, \bi_{h} \}$ which is a violating triangle in $M$, output ``reject.'' Otherwise, output ``accept.''
    \end{enumerate}
\end{framed}

\begin{lemma}[$\CheckHiDegree$ Lemma]\label{lem:check-hi-degree}
For $n \in \N$ and $\eps \in (1/n,1)$, there exists a randomized algorithm, $\CheckHiDegree$, which receives as input an $n\times n$ matrix $M \in \calC$ and a parameter $\eps$ and has the following guarantees:
\begin{itemize}
\item If $M \in \calP$, $\CheckHiDegree(M,\eps)$ always outputs ``accept.'' 
\item Letting $T$ be the set of violating triangles of $M$, if there are at least $\eps n/\sco$ indices $i \in [n]$ such that $d_{T}(i) \geq \ea n^{4/3}/16$, $\CheckHiDegree(M, \eps)$ outputs ``reject'' with probability at least $5/6$. 
\end{itemize} 
The algorithm is non-adaptive, taking $O(1/\eps + n^{2/3}/\ea)$ samples and using $O(n^{2/3} / \eps^{4/3})$ queries.
\end{lemma}
\begin{proof}
    See appendix.
\end{proof}

\begin{lemma}[$\CheckViolation$ Lemma]\label{lem:check-vio}
For $n \in \N$ and $\eps \in (1/n,1)$, there exists a randomized algorithm, $\CheckViolation$, which receives as input an $n\times n$ matrix $M \in \calC$ and a parameter $\eps$ and has the following guarantees:
\begin{itemize}
\item If $M \in \calP$, $\CheckViolation(M,\eps)$ always outputs ``accept.''
\item Letting $T$ be the set of violating triangles of $M$, if $M$ is $\eps$-far from $\calP$ and the set of indices $i \in [n]$ with $d_T(i) \geq \ea n^{4/3}/16$ has size at most $\eps n/\sco$, the sub-routine outputs ``reject'' with probability at least $5/6$.
\end{itemize}
The algorithm is non-adaptive, taking $O(n^{1/3} / {\eps^{2/3}})$ samples and using $O(n^{2/3} /{\eps^{4/3}})$ queries.
\end{lemma}
\begin{proof}[Proof of Theorem~\ref{thm:testing-metric} Assuming Lemma~\ref{lem:check-vio}] 
See appendix.
\end{proof}

\subsection{CheckViolation Sub-routine: Proof of Lemma~\ref{lem:check-vio}}

Note that the sub-routine $\CheckViolation$ only outputs ``reject'' when it observes a violating triangle. Therefore, it is easy to establish the first condition of Lemma~\ref{lem:check-vio}, as there are no violating triangles whenever $M \in \calP$. It remains to prove that, whenever $M$ is $\eps$-far from $\calP$ and at most $\eps n/4$ indices $i \in [n]$ have $d_T(i) \geq \ea n^{4/3}/16$, the sub-routine finds a violation with probability at least $5/6$. 
\begin{Definition}
For $M \in \calC$ and any subset $\tilde{T} \subseteq T$ of violating triangles of $M$, the random variable $\bX(\tilde{T}) \geq 0$ is given by
\[ \bX(\tilde{T}) = \sum_{t \in \tilde{T}} \ind\left\{ t \subset \{ \bi_1,\dots, \bi_s \} \right\}. \]
where $\bi_1,\dots, \bi_s \sim [n]$ are indices sampled from $\CheckViolation$.
\end{Definition}

\begin{claim}\label{cl:exp-var}
    For any collection of triangles $\tilde{T}$, the expectation of $\bX(\tilde{T})$ is bounded by
    \begin{align*}
    \Ex_{\bi_1,\dots, \bi_s}\left[ \bX(\tilde{T}) \right] \geq \Omega\left( \frac{|\tilde{T}| \cdot s^3}{n^3} \right),
    \end{align*}
    whenever $3 \leq s \ll n$. Moreover, the variance of the random variable is bounded by 
\[
\Varx_{\bi_1,\dots, \bi_s}\left[ \bX(\tilde{T}) \right] \lsim \Ex_{\bi_1,\dots, \bi_s}\left[ \bX(\tilde{T})\right] + \left(\sum_{i \in [n]} d_{\tilde{T}}(i)^2 \right) \cdot \dfrac{s^5}{n^5} + \left( \sum_{i \neq j} d_{\tilde{T}}(i,j)^2 \right) \cdot \frac{s^4}{n^4},
\]
where the second term counts pairs of triangles $t, t' \in \tilde{T}$ which intersect at a single vertex, and the third term counts the number of pairs of triangles $t, t' \in \tilde{T}$ which intersect at two vertices.
\end{claim}
\begin{proof}
    By linearity of expectation, \begin{align*}
    \Ex_{\bi_1,\dots, \bi_s}\left[ \bX(\tilde{T}) \right] &\geq |\tilde{T}| \sum_{1 \leq k_1 < k_2 < k_3 \leq s} 3! \cdot \Prx\left[ \begin{array}{c} \bi_{k_1} = 1, \bi_{k_2} = 2, \bi_{k_3} = 3 \\ \text{and occur uniquely} \end{array} \right] \\
        &\geq |\tilde{T}| \cdot \binom{s}{3} \cdot 3! \cdot \frac{1}{n^3} \left(1 - \frac{3}{n} \right)^{s-3} = \Omega\left( \frac{|\tilde{T}| \cdot s^3}{n^3} \right).
    \end{align*}
    On the other hand,
    \begin{align*}
\Varx_{\bi_1,\dots, \bi_s}\left[ \bX(\tilde{T}) \right] &= \Ex_{\bi_1,\dots, \bi_s}\left[ \left( \sum_{t \in \tilde{T}} \ind\left\{ t \subset \{ \bi_1,\dots, \bi_{s} \} \right\} \right)^2 \right] - \Ex_{\bi_1,\dots, \bi_s}\left[ \sum_{t \in \tilde{T}} \ind\left\{ t \subset \{ \bi_1,\dots, \bi_{s} \} \right\} \right]^2 \\
			&\lsim \Ex_{\bi_1,\dots, \bi_s}\left[ \bX(\tilde{T})\right] + \left(\sum_{i \in [n]} d_{\tilde{T}}(i)^2 \right) \cdot \dfrac{s^5}{n^5} + \left( \sum_{i \neq j} d_{\tilde{T}}(i,j)^2 \right) \cdot \frac{s^4}{n^4}.
\end{align*}
\end{proof}

Since $\bX(\tilde{T})$ counts the number of violating triangles from $\tilde{T}$ included in the random sample $\bi_1,\dots, \bi_s$, if $\bX(\tilde{T}) > 0$, then the algorithm has sampled a violating triangle. Once it makes all pairwise queries, it will find the violating triangle and output ``reject.'' The analysis will find a subset $\tilde{T}$ of violating triangles such that the random variable $\bX(\tilde{T})$ is non-zero with high constant probability, which indicates that the random sample contains at least one violating triangle. More specifically, the goal is to find an appropriate set of violating triangles $\tilde{T}$ such that the expectation of $\bX(\tilde{T})$ is large while the variance is small. Hence, the $\tilde{T}$ used in the analysis has large cardinality, and both vertex-triangle-degree and edge-triangle-degree of the indices defined over $\tilde{T}$ are bounded.

\begin{lemma}\label{lem:goodSet}
    Let $M\in\calC$ be $\eps$-far from $\calP$ and let $T$ denote the set of all violating triangles of $M$. Suppose that the set of indices $i\in[n]$ such that $d_T(i)\geq \ea n^{4/3}/16$ has size at most $\eps n/\sco$. Then, there exists a subset $\tilde{T}$ of violating triangles of $M$ satisfying
    \begin{itemize}
        \item $|\tilde{T}|\geq \Omega(\eps n^2),$
        \item For every index $i\in[n],d_{\tilde{T}}(i)\leq O(\ea n^{4/3}),$
        \item for every pair of indices $(i,j)\in [n]\times [n]$, $d_{\tilde{T}}(i,j)\leq O( n^{2/3}/\eps^{1/3})$.
    \end{itemize}
\end{lemma}

\begin{proof}[Proof of Lemma~\ref{lem:check-vio} assuming Lemma~\ref{lem:goodSet}]
    The sub-routine $\CheckViolation(M,\eps)$ samples $c\cdot n^{1/3}/\eps^{{2/3}}$ indices uniformly at random for some constant $c$. Since it is guaranteed that there are at least $\Omega(\eps n^2)$ violating triangles in the set $\tilde{T}$, Claim~\ref{cl:exp-var} implies the expected value of $\bX(\tilde{T})$ is a large constant for $c$ large enough. From Lemma~\ref{lem:goodSet}, we derive the following two inequalities:
    \begin{align*}
    \sum_{i \in [n]} d_{\tilde{T}}(i)^2 &\leq O(\ea n^{4/3}) \sum_{i \in [n]} d_{\tilde{T}}(i) = O\left(\ea n^{4/3} \cdot |\tilde{T}|\right), \\
    \sum_{(i,j) \in [n] \times [n]} d_{\tilde{T}}(i,j)^2 &\leq O( n^{2/3}/\eps^{1/3}) \sum_{(i,j) \in [n]\times [n]} d_{\tilde{T}}(i,j) = O\left(  n^{2/3}/{\eps^{1/3}} \cdot |\tilde{T}|\right).
    \end{align*}
Consider (for the sake of analysis) repeating the above randomized trial for $\bX(\tilde{T})$ for $k$ independent iterations, letting $\bX_1(\tilde{T}), \dots, \bX_k(\tilde{T})$ denote the outcomes of $k$ independent trials. Then, taking the average and applying Chebyshev's inequality,
\begin{align*}
\Prx\left[ \frac{1}{k} \sum_{\ell=1}^k \bX_{\ell}(\tilde{T}) = 0\right] \leq \dfrac{\Var[\bX(\tilde{T})]}{k \cdot \Ex[\bX(\tilde{T})]^2} &\lsim \frac{\Ex[\bX(\tilde{T})]}{k\cdot \Ex[\bX(\tilde{T})]^2} + \frac{\ea n^{4/3} \cdot |\tilde{T}| \cdot \frac{s^5}{n^5}}{k \cdot \Ex[\bX(\tilde{T})]^2} + \frac{ n^{2/3}/\eps^{1/3} \cdot |\tilde{T}| \cdot \frac{s^4}{n^4}}{k\cdot  \Ex[\bX(\tilde{T})]^2} \\
    &= O\left(\frac{n^3}{k \cdot |\tilde{T}| \cdot s^3} \right) + O\left(\frac{\ea n^{7/3}}{k \cdot |\tilde{T}| \cdot s} \right) + O\left( \frac{n^{8/3}}{\eps^{1/3}\cdot k \cdot |\tilde{T}| \cdot s^2} \right),
\end{align*}
which can be made an arbitrarily small constant when $|\tilde{T}| = \Omega(\eps n^2)$, $s = O(n^{1/3}/{\eps^{2/3}})$ and $k = O(1)$. Notice that the second and the third terms are the bottle-neck.

\end{proof}

\subsection{Finding A Good Subset of Violating Triangles: Proof of Lemma~\ref{lem:goodSet}}

\begin{lemma}\label{lem:vertex-deg}
    Suppose $M\in \calC$ and is $\eps$-far from $\calP$ and there are at most ${\eps n/\sco}$ indices $i\in[n]$ such that the vertex-triangle degree $d_{T}(i)\geq \ea n^{4/3}/16$. Then there exists a matrix $M'\in\calC$ which is $\eps/2$-far from $\calP$ such that all violating triangles in $M'$ are also in $M$. Denote this set of violating triangles in $M'$ to be $T'$. Moreover, for all $i\in[n]$, $d_{T'}(i)< \ea n^{4/3}/16$. 
\end{lemma}
\begin{proof}
    Let $m \in \R$ be the value of the maximum entry in $M$, and let $I \subset [n]$ be the set of indices $i \in [n]$ with $d_T(i) \geq \ea n^{4/3}/16$. We consider the following $n\times n$ matrix $M'$, where we set $M'(i,j) = M(i,j)$ unless, either $i$ or $j$ lie in $I$, in which case $M'(i,j) = m$. Note that $\| M' - M \|_0 \leq 2 |I| \cdot n \leq 2\eps n^2/\sco = \eps n^2/2$, so that $M'$ is $\eps/2$-far from $\calP$. We apply Lemma~\ref{lem:exist} to $M'$, which shows that there are at least $\eps n^2/{12}$ distinct violating triangles $\{i,j,k\}$ in $M'$. Note that, for any violating triangle $\{i,j,k\}$ in $M'$, $i,j,k$ are not in $I$ otherwise $\{i,j,k\}$ is not violating in $M'$. Hence, the distances between these three vertices are not changed, so this triangle is also violating in $M$. This shows $T'$, the set of all violating triangles in $M'$, is a subset of $T$. Thus, for $i\notin I$, $d_{T'}(i)\leq d_T(i)<\ea n^{4/3}/16$. For $i\in I$, no violating triangle in $T'$ contains $i$, so $d_{T'}(i)=0$.
\end{proof}

\begin{Definition}
    Let the set of ``high-degree-edges" $\overline{E}$ to be the set of pairs of indices $(i,j)$ such that $d_{T'}(i,j)>10 n^{2/3}/{\eps^{1/3}}$.
\end{Definition}
\begin{lemma}\label{lem:low-heavy-edge}
    For any $i\in[n],$ the number of edges in $\overline{E}$ that contain $i$ is upper bounded by ${\eps^{2/3}} n^{2/3}{/80}$. Therefore, the size of $\overline{E}$ is upper bounded by ${\eps^{2/3}n^{5/3}/80},$ which is at most $\eps n^2/80$ for $\eps>1/n$.
\end{lemma}
\begin{proof}

    Notice that $d_{T'}(i)=\sum_{e:i\in e}d_{T'}(e)/2\geq 10 n^{2/3}/{{\eps^{1/3}}}\cdot|\{e\in \overline{E}:i\in e\}|/2$. By the bound on the vertex-triangle-degree defined over $T'$ given in Lemma~\ref{lem:vertex-deg}, $d_{T'}(i)\leq \ea n^{4/3}/16$. Thus, $|\{e\in \overline{E}:i\in e\}|\leq {\eps^{2/3}} n^{2/3}/80$.
\end{proof}

\begin{Definition}[Unique triangles]
    For each edge $e\in\overline{E}$, define the unique triangles to $e$ to be the subset $T_e^{(u)}\subset T'$ such that each triangle in $T_e^{(u)}$ only uses edge $e$ and nothing else in $\overline{E}$.
\end{Definition}

\begin{lemma}\label{lem:many-unique}
    For any edge $e\in \overline{E}$, $|T_e^{(u)}|\geq 9 n^{2/3}/{{\eps^{1/3}}}$.
\end{lemma}
\begin{proof}
    Let edge $e=(i,j)$. Since $e\in \overline{E},$ $e$ is a high-degree edge: $d_{T'}(e)>10 n^{2/3}/{{\eps^{1/3}}}$. By Lemma~\ref{lem:low-heavy-edge}, we can upper bound the number of violating triangles that use $e$ and some other high-degree edge $(i,v)\in\overline{E}$ since few high-degree edges are using the vertex $i$. Quantitatively, we have
    \[
     \left|\Bigl\{\{i,j,v\}\in T':(i,v)\in \overline{E}\Bigr\}\right|
     \leq 
     \left|\Bigl\{\{i,j,v\}\subset [n]:(i,v)\in \overline{E}\Bigr\}\right|\leq {\eps^{2/3}} n^{2/3}/80.
    \]
    And similarly for vertex $j$, we have
    \[
    \left|\Bigl\{\{i,j,v\}\in T':(j,v)\in \overline{E}\Bigr\}\right|
    \leq
    \left|\Bigl\{\{i,j,v\}\subset [n]:(j,v)\in \overline{E}\Bigr\}\right|\leq {\eps^{2/3}} n^{2/3}/80.
    \]
    Therefore, the violating triangles that remain and only use edge $e$ can be bounded below.
    \[
    |T_e^{(u)}|=\left|\Bigl\{\{i,j,v\}\in T':(i,v),(j,v)\notin \overline{E}\Bigr\}\right|\geq d_{T'}(i,j)-\eps^{2/3}n^{2/3}/40\geq9 n^{2/3}/{{\eps^{1/3}}}.
    \]
\end{proof}

Notice that Lemma~\ref{lem:vertex-deg} above guarantees the existence of a subset $T'\subset T$ of violating triangles such that the vertex-triangle-degree defined over $T'$ at each vertex is bounded above. The construction of $T'$ is through modifying all entries, or edges, that previously contained a high-degree vertex. The following lemma continues the modification and constructs a subset of violating triangles $\fT\subset T'$ such that the \textit{edge-triangle-degree }defined over $\fT$ at each edge is bounded above. The idea is to modify all high-degree entries, or edges. Recall that $\overline{E}$ denotes the set of high-degree edges. 

\begin{lemma}\label{lem:modify}
    Suppose matrix $M'$ and the set of violating triangles $T'$ are as defined in Lemma~\ref{lem:vertex-deg}. Then there exists a matrix $M''$ and a subset of violating triangles $\fT\subset T'$ that satisfy the below properties:
    \begin{itemize}
        \item $M'$ and $M''$ differ on only the entries specified in $\overline{E}$
        \item Suppose $T''$ is the set of all violating triangles in $M''$, then $\fT\subset T'$ has size at least $|T''|$
        \item For all $i,j\in[n],$ $d_{\fT}(i,j)\leq 10 n^{2/3}/{{\eps^{1/3}}}$.
    \end{itemize}
\end{lemma}
\begin{proof}[Proof of Lemma~\ref{lem:goodSet} assuming Lemma~\ref{lem:modify}]
By Lemma~\ref{lem:low-heavy-edge}, $|\overline{E}|\leq \eps n^2/80$. Moreover, by construction in Lemma~\ref{lem:vertex-deg}, $M'$ and $M$ differ on at most $\eps n^2/2$ entries. Since $M'$ and $M''$ differ on only $2|\overline{E}|\leq \eps n^2/40$ entries, $M$ and $M''$ differ on at most $21\eps n^2/40$ entries, meaning that $M''$ is $19\eps/40$-far from $\calP$. Again, using Lemma~\ref{lem:exist}, there are at least $19\eps n^2/240$ violating triangles in $M''$. That is, $|T''|\geq 19\eps n^2/240$. Lemma~\ref{lem:modify} promises the existence of a set of violating triangles $\fT\subset T'$ with size at least $|T''|\geq 19\eps n^2/240$, and $d_{\fT}(i,j)\leq 10n^{2/3}/{{\eps^{1/3}}}$ for all $i,j\in[n]$. Moreover, Lemma~\ref{lem:vertex-deg} states that every triangle in $T'$ is violating in $M$. Thus, $\fT$ is a subset of violating triangles in $M$. Lastly, for any index $i\in[n]$, $d_{\fT}(i)\leq d_{T'}(i)\leq \ea n^{4/3}/16$.
\end{proof}
\begin{proof}[Proof of Lemma~\ref{lem:modify}]
    First, we specify the values $M''$ takes on for each entry in $\overline{E}$ (Claim~\ref{lem:mod-val}). Then, we provide a mapping from $T''$ to a subset of violating triangles $\fT\subset T'$ in $M'$ (Claim~\ref{lem:shuffle_tri}) and show that $\fT$ has the claimed properties.

    Recall that a high-degree edge in $\overline{E}$ corresponds to an entry in $M'$ that participates in many violating triangles. The claim below (Claim~\ref{lem:mod-val}) states that there exists some new value $x \in \R_{\geq 0}$ such that, if the entry $e \in \ol{E}$ of the matrix $M$ is changed to the new value, the entry participates in much fewer violating triangles in the new matrix. We give a notation for finding a new value for a high-degree edge. For indices $i,j,k\in[n]$ and for any value $x\in\mathbb{R}_{>0}$, let $\mathbb{I}(i,j,k,x)$ denote the indicator variable in $\{0,1\}$ with
    \[
    \mathbb{I}(i,j,k,x)= 1 \qquad \iff \qquad \Biggl\{
    \begin{array}{c}
    (i,k)\notin\overline{E},\\
    (j,k)\notin\overline{E},\\
    x\notin[|M'(i,k)-M'(j,k)|,M'(i,k)+M'(j,k)]
    \end{array}
    \Biggr\}.
    \]

\begin{claim}\label{lem:mod-val}
    For each $e=(i,j)\in\overline{E}$, whenever $d_{T'}(i)$ and $d_{T'}(j)$ are both bounded above by $\ea n^{4/3}/16$, there exists a value $x(e) \in \R_{\geq 0}$ such that 
    \[ \sum_{k \in [n]} \mathbb{I}(i,j,k, x(e)) \leq \eps^{1/6} \cdot n^{2/3}.\]
\end{claim}

Assuming Claim~\ref{lem:mod-val} (which we formally prove shortly), let
\[ M''(i,j) = \left\{\begin{array}{cc} M'(i,j) & (i,j) \notin \ol{E} \\
    x(e) & (i,j) = e \in \ol{E} \end{array} \right. ,\]
as the setting $x(e)$ from Claim~\ref{lem:mod-val}, and let $T''$ denote all violating triangles of $M''$.

\begin{claim}
    For every edge $e=(i,j)\in\overline{E}$, let $A_e$ be the subset of triangles in $T''$ that contain the edge $e$. Then, $d_{T''}(e)=|A_e|\leq 2{\eps^{1/6}}n^{2/3}$.
\end{claim}
\begin{proof}
    Following Claim~\ref{lem:mod-val}, we can upper bound the number of triangles in $T''$ which only use edge $e$ and nothing else in $\overline{E}.$ That is,
\[
\left|\Bigl\{\{i,j,k\}\in T'':(i,k),(j,k)\notin \overline{E}\Bigr\}\right|=\sum_{k\in[n]}\mathbb{I}(i,j,k,x(e))\leq {\eps^{1/6}}n^{2/3}.
\]
On the other hand, few triangles use $e$ and some other edge in $\overline{E}$, as implied by Lemma~\ref{lem:low-heavy-edge}. That is,
\[
\left|\Bigl\{\{i,j,k\}\in[n]:(i,k)\text{ or }(j,k)\in \overline{E}\Bigr\}\right|\leq {\eps^{2/3}} n^{2/3}/40.
\]
Together, these two sets of triangles consist all of $A_e$. Hence, the edge-triangle degree of $e$ over the set $T''$ is upper bounded by $d_{T''}(e)=|A_e|\leq \eps^{1/6}n^{2/3}+\eps^{2/3}n^{2/3}/40\leq2{\eps^{1/6}}n^{2/3}$. 
\end{proof}
With the set of all violating triangles $T''$ in $M''$, we then use the below helper lemma to construct a subset of violating triangles $\fT\subset T'$ that inherits the nice property of $T''$: the edge-triangle-degree defined over $\fT$ for each edge $e$ is upper bounded.
\begin{claim}\label{lem:shuffle_tri}
    For each $e=(i,j)\in\overline{E}$, let $A_e$ be the subset of triangles in $T''$ that contain the edge $e$. Then, there exists a subset of violating triangles $B_e\subset T_e^{(u)}\subset T'$ such that $|A_e|=|B_e|$.
\end{claim}
\begin{proof}[Proof of Claim~\ref{lem:shuffle_tri}]
    Notice that $|A_e|=d_{T''}(e)\leq 2{\eps^{1/6}}n^{2/3}$. By Lemma~\ref{lem:many-unique}, $|T_e^{(u)}|\geq 9n^{2/3}/{{\eps^{1/3}}}$. This suggests the existence of $B_e\subset T_e^{(u)}\backslash A_e$ with size $|A_e|\leq 2{\eps^{1/6}}n^{2/3}$.
\end{proof}

Claim~\ref{lem:shuffle_tri} above hints on how to construct a mapping from $T''$ to $\fT$. We initialize an empty set $\fT$. First, add to $\fT$ all the triangles in $T''$ that do not use any of the high-degree edges in $\overline{E}$. Then, for each $e\in\overline{E}$, following Claim~\ref{lem:shuffle_tri}, find the subset $B_e\subset T_e^{(u)}$ and add $B_e$ to $\fT$. What we are left to show is that $\fT$ has the properties claimed in Lemma~\ref{lem:modify}. That is, $|\fT|\geq |T''|$ and for any edge $(i,j)\in[n]\times[n],$ $d_{\fT}(i,j)\leq 10n^{2/3}/{{\eps^{1/3}}}$.

\begin{claim}
    Define $\fT$ to be \[
    (\bigcup_{e\in\overline{E}} B_e)\cup \{\{i,j,k\}\in T'':(i,j),(j,k),(i,k)\notin\overline{E}\}.
    \] Then,
    $|\fT|\geq |T''|$. Furthermore, $\fT\subset T'$.
\end{claim}
\begin{proof}
    The inequality follows from the fact that each triangle in $ T''$ corresponds to at least one unique triangle in $\fT$. Notice that 
    \[
    T''= (\bigcup_{e\in\overline{E}} A_e)\cup \{\{i,j,k\}\in T'':(i,j),(j,k),(i,k)\notin\overline{E}\}.
    \]
    The set of triangles that do not use any high-degree edge $\overline{E}$ is exactly the same in $T''$ and $\fT$. On the other hand, notice that each $B_e$ is disjoint from the set $\{\{i,j,k\}:(i,j),(j,k),(i,k)\notin\overline{E}\}$. Moreover, $B_e$ and $B_{e'}$ are disjoint whenever $e\neq e'$ since $B_e$ is a subset of $T_e^{(u)}$. Since $|A_e|=|B_e|,$ for each edge $e\in\overline{E}$ there is some bijective map from $A_e$ to $B_e$. Thus, each triangle in $T''$ corresponds to at least one unique triangle in $\fT$, which suggests $|\fT|\geq |T''|$. Note that the inequality may be strict because $A_e$ might overlap with $A_{e'}$ for $e\neq e'$, and it only makes the union larger: $|B_e\cup B_{e'}|>|A_e\cup A_{e'}|$.

    It is clear that every triangle in $B_e$ for each $e$ is in $T_e^{(u)}\subset T'$. Moreover, each triangle in $\{\{i,j,k\}\in T'':(i,j),(j,k),(i,k)\notin\overline{E}\}$ has same distance values on all three edges in $M'$ and $M''$. Since they are violating in $M''$, they are also violating in $M'$, which indicates $\{\{i,j,k\}\in T'':(i,j),(j,k),(i,k)\notin\overline{E}\}\subset T'$.
\end{proof}

\begin{claim}
    For every edge $e=(i,j)$, $d_{\fT}(e)\leq 10n^{2/3}/{\eps^{1/3}}.$
\end{claim}
\begin{proof}
    If $e\notin\overline{E}$, by definition $d_{\fT}(e)\leq d_{T'}(e)\leq 10n^{2/3}/{{\eps^{1/3}}}.$ If $e\in\overline{E}$, the only triangles in $\fT$ that contain $e$ are the triangles in $B_e$. Hence, $d_{\fT}(e)=|B_e|=|A_e|=d_{T''}(e)\leq 10n^{2/3}/{{\eps^{1/3}}}.$
\end{proof}
The two above Claims complete the proof for Lemma~\ref{lem:modify}, and we are left to show the correctness of Claim~\ref{lem:mod-val}.

\begin{proof}[Proof of Claim~\ref{lem:mod-val}]
    The statement follows from the below claim.
    \begin{claim}\label{c:intervals}
        Suppose for the edge $e=(i,j)\in\overline{E}$ and any value $x(e)\in(0,\infty)$, $\sum_{k\in[n]}\mathbb{I}(i,j,k,x(e))\geq t$, then either $d_{T'}(i)\geq t^2/8$ or $d_{T'}(j)\geq t^2/8$.
    \end{claim}
    Suppose Claim~\ref{c:intervals} holds. We take $t$ to be ${\eps^{1/6}}n^{2/3}$. If for any value $x(e)\in(0,\infty)$, $\sum_{k\in[n]}\mathbb{I}(i,j,k,x(e))\geq t={\eps^{1/6}}n^{2/3}$, then we have a contradiction since the vertex-triangle-degrees of both $i$ and $j$ defined over $T'$ are at most $\ea n^{4/3}/16$. Thus, we are done if the above claim holds.
    \begin{proof}[Proof of Claim~\ref{c:intervals}]
        First, the indicator variable $\mathbb{I}(i,j,k,x(e))$ equals 1 if and only if both edges $(i,k),(j,k)\notin\overline{E}$ and the value $x(e)$ does not lie within the interval $[|M'(i,k)-M'(j,k)|,M'(i,k)+M'(j,k)]$. 
        We will then show that if for all $x(e) \in (0,\infty)$, 
        \[ \sum_{k \in [n]} \mathbb{I}(i,j,k,x(e)) \geq t,\]
        then there exist at least $t^2/4$ pairs $(k_1,k_2) \in [n] \times [n]$ such that the intersection of intervals 
        \[ \left[ |M'(i,k_1) - M'(j,k_1)|, M'(i,k_1) + M'(j,k_1) \right] \cap \left[ |M'(i, k_2) - M'(j,k_2)|, M'(i,k_2) + M'(j,k_2) \right] = \emptyset.\]
        The statement is true more generally, if there is a collection of $m$ intervals $I_1,\dots, I_m$ of the real line, and for all $x$, there are at least $t$ intervals which do not contain $x$, then there are two subsets $L$ and $R$ of at least $t/2$ intervals where $l \in L$ and $r \in R$ are disjoint, giving us $t^2/4$ pairs. The proof is algorithmic:
        \begin{itemize}
            \item Consider sorting the intervals by increasing endpoints, and we will let $x$ scan from $-\infty$ to $\infty$ according to the endpoints of intervals. Let $x_{t/2}$ denote the endpoint of the $t/2$-th interval and let $D$ denote the set of intervals disjoint from $x_{t/2}$. 
            \item Note that, there must be at most $t/2$ intervals which lie to the left of $x_{t/2}$, since we scanned in increasing order of endpoint; and since $D$ is disjoint, there must be at least $t - (t/2) \geq t/2$ intervals whose start is larger than $x_{t/2}$. Thus, we let $L$ denote the $t/2$ intervals of first endpoints, and we let $R$ denote the intervals in $D$ whose start points come after $x_{t/2}$. 
        \end{itemize}
        

        Now, consider the implication that two intervals $[|M'(i,k_1)-M'(j,k_1)|,M'(i,k_1)+M'(j,k_1)]$ and $[|M'(i,k_2)-M'(j,k_2)|,M'(i,k_2)+M'(j,k_2)]$ are disjoint, for two vertices $k_1,k_2$. Without loss of generality, we assume that $|M'(i,k_1)-M'(j,k_1)|\leq M'(i,k_1)+M'(j,k_1)<|M'(i,k_2)-M'(j,k_2)|\leq M'(i,k_2)+M'(j,k_2)$. Then, from the middle inequality, it must be the case that either $M'(i,k_2)>M'(i,k_1)+M'(j,k_1)+M'(j,k_2)$ or $M'(j,k_2)>M'(i,k_1)+M'(j,k_1)+M'(i,k_2)$. Intuitively, it means the 4-cycle $\{i,k_1,j,k_2\}$ is ``violating". Suppose we are in the case that $M'(i,k_2)$ is larger than the sum of the other three edges. (The other case that $M'(j,k_2)$ is the largest follows the same analysis.) Then, it must be the case that either $M'(i,k_2)>M'(i,k_1)+M'(k_1,k_2)$ or $M'(k_1,k_2)>M'(j,k_1)+M'(j,k_2)$. In the former case, edge $(k_1,k_2)$ forms a violating triangle with index $i$ while in the latter case, edge $(k_1,k_2)$ forms a violating triangle with index $j$.

        Combining all the above arguments, there are at least $t^2/4$ pairs of pairwise disjoint intervals. Each pair corresponds to a unique edge $(k_1,k_2)$ for some $k_1,k_2\in[n]$ which forms a violating triangle with either $i$ or $j$. By pigeon-hole theorem, at least $t^2/8$ distinct edges form some violating triangles with either $i$ or $j$, implying that $d_{T'}(i)\geq t^2/8$ or $d_{T'}(j)\geq t^2/8$. The claim follows.
\end{proof}

\end{proof}

\end{proof}

\section{Metric Testing Lower Bound---Theorem~\ref{thm:testing-metric-lb}}

We specify a distribution $\Dno$ which is supported on $(3n) \times (3n)$ matrices that are $\eps$-far from being a metric, where $\eps = n^{-\nu(n)}$, for $\nu(n) = (\log \log \log n + 4) / \log \log n$. We show that a non-adaptive algorithm which makes $o(n^{2/3 + 2\nu(n)/3})$ queries will not be able to find a violation of the triangle inequality with high probability when the input $\bM \sim \Dno$. The construction follows closely to the Behrend graph construction used in~\cite{AKKR08} for showing the hardness of testing triangle-freeness. In particular, we make use of a Salem-Spencer set, which is a dense set of numbers that is free of $3$-arithmetic progressions.

\begin{lemma}[Lemma~4 in~\cite{AKKR08}]\label{lem:ap-free}
    For any sufficiently large $n$, there exists a set $X \subset [n]$ of size at least $n^{1-\nu(n)}$, for $\nu(n) = (\log \log \log n + 4) / \log \log n$, so that for all $x, y, z \in X$, $x + y \equiv 2z \mod n$ if and only if $x = y = z$.
\end{lemma}

\paragraph{Description of $\Dno$.} We describe a single $(3n) \times (3n)$ matrix $M$, which we will then randomize by re-shuffling indices in $[3n]$. Our construction follows closely that of the Behrend graph (Section 5.3 in~\cite{AKKR08}), which is a tripartite graph on $3n$ vertices (each part with $n$ vertices) and $n|X|$ edge-disjoint triangles, where set $X$ is from Lemma~\ref{lem:ap-free}. For simplicity, we partition the set of $3n$ indices into $A$, $B$, and $C$, each of size $n$, and we will associate each $a \in A$ with a number in $[n]$. Hence, we refer to $M(a, b)$ as the entry in the matrix $M$ of row $a \in A$ and column $b \in B$---this causes some ambiguity among the indices, since $a\in A$ both corresponds to a number, and the fact we use ``$a$'' means we refer to the subset of rows/columns associated with $A$ (similarly for $b$ and $c$). We let:
\begin{itemize}
    \item For any $a \in A$ and $b \in B$, we let $M(a,b) = M(b,a) = 1$ if and only if $b- a \equiv x \mod n$ with $x \in X$, and we let $M(a,b) = M(b,a) = 2$ otherwise. 
    \item For any $b \in B$ and $c \in C$, we let $M(b, c) = M(c,b) = 2$ if and only if $c - b \equiv x \mod n$ with $x \in X$, and we let $M(b,c) = M(c,b) = 3$ otherwise.
    \item For any $a \in A$ and $c \in C$, we let $M(a, c) = M(c, a) = 4$ if and only if $c - a \equiv 2x \mod n$ with $x \in X$ and $M(a, c) = M(c, a) = 2$ otherwise.
    \item Every other non-diagonal entry $M(x,y) = 2$ and every diagonal entry $M(x,x) = 0$.
\end{itemize}
We consider the following important properties of the construction above:
\begin{claim}\label{cl:matrix-construction}
For the matrix $M$ constructed above, we have:
\begin{itemize}
\item Suppose $i, j, k$ are three indices that form a violating triangle. Then, up to a permutation, $M(i,j) = 1$, $M(j,k) = 2$, and $M(k, i) = 4$.
\item For the permutation where the above holds, there is a single entry $i = a \in A$, $j = b \in B$, and $k = c \in C$.
\item A violating triangle $\{a, b, c\}$ as above is uniquely associated with $a \in A$ and $x \in X$, so the violating triangle consists of $\{ a, a + x\mod n, a + 2x \mod n\}$. 
\end{itemize}
The above implies that $2n|X|$ modifications on the entries of $M$ are required in order to remove all violations of the triangle inequality. Furthermore, the guarantee of Lemma~\ref{lem:ap-free} ensures that there are exactly $n|X|$ triples $\{a,b,c\}$ which violate the triangle inequality.
\end{claim}

\begin{figure}\label{fig:hard-behrend}
    \centering
    \begin{picture}(200, 200)
        \put(0,0){\includegraphics[width=0.5\linewidth]{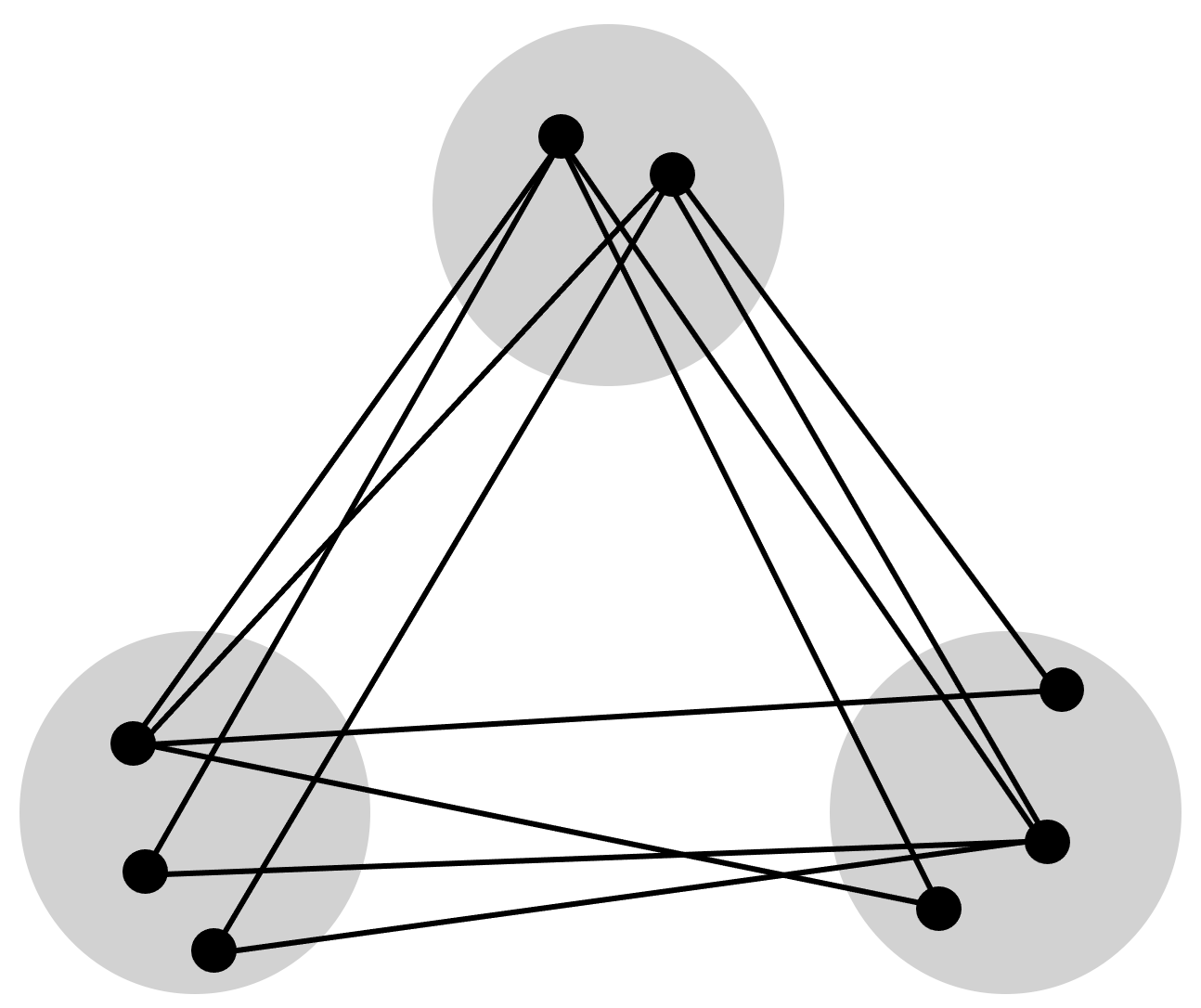}}
        \put(70, 170){$A$}
        \put(-10, 50){$B$}
        \put(235, 50){$C$}
        \put(60, 120){$1$}
        \put(120, 10){$2$}
        \put(180, 120){$4$}
    \end{picture}
\end{figure}

\begin{proof}[Proof of Claim~\ref{cl:matrix-construction}] First, notice that all distances are $\{0,1,2,3,4\},$ where distances of $0$ are solely for $M(x,x)$. Thus, violations of the triangle inequality come from the possible tuples $(1,1,3),(1,1,4),(1,2,4)$. However, we note that one cannot have violations of the triangle inequality of the form $(1, 1, 3)$ or $(1, 1, 4)$. The reason is that $M(i,j) = 1$ occurs only if indices $i, j$ correspond to some $a \in A$ and $b \in B$. Thus, if $M(i,j) = 1$ and $M(j,k) = 1$, then $i,j,k$ are all among $A\cup B$. However, the only entries set to $3$ or $4$ are incident on $C$. 

Furthermore, violations of the form $(1, 2, 4)$ are exactly those arising from the $n|X|$ edge-disjoint triangles $\{a, b, c\}$ specified by some $x \in X$, where $b - a \equiv x \mod n$, $c - b \equiv x \mod n$ and $c- a \equiv 2x \mod n$. This follows from the fact that $X$ is three of length-$3$ arithmetic progressions. Since these violating triangles are edge-disjoint, each entry in $M$ is involved in at most one 3-tuples violating the triangle inequality. Moreover, since at least one entry in $M(i,j), M(i,k), M(j,k)$ has to be modified for the triple $\{i,j,k\}$ to obey metric property, we need to modify at least one edge in each violating triple. That is, at least $n|X|$ entries in $M$ has to be modified for $M$ to be a metric. In fact, exactly $n|X|$ modifications suffice: simply modify all edges with weight 4 to 3. There are $n|X|$ such edges in $G$, which correspond to $2n|X|$ such entries in $M$.
\end{proof}

See Figure~\ref{fig:hard-behrend} for a description of the above construction, see also~Section~2.6 in~\cite{Z23b}. We let $\bM \sim \Dno$ be obtained from $M$ by re-ordering rows and columns according to a uniformly random permutation $\bpi$. By Yao's minimax principle, in order to rule out non-adaptive algorithms that have one-sided error (i.e., always accept metrics), it suffices to rule out any deterministic and non-adaptive algorithm for finding violating triangles in a draw $\bM \sim \Dno$.

\begin{lemma}\label{lem:algo-lb-1}
    Consider any deterministic non-adaptive algorithm that makes $o(n^{2/3+2\nu(n)/3})$ queries. With probability at least $2/3$ over the draw of $\bM \sim \Dno$, there are no queries $(i,j), (j,k),$ and $(i,k)$ with 
    \[ \bM(i,j) > \bM(i,k) + \bM(k,j).\]
    Since any partial matrix that does not violate the triangle inequality may be completed to one which is a metric (by considering the observed weighted sub-graph and computing shortest paths, as well as a maximum distance), any non-adaptive and one-sided algorithm must make $\Omega(n^{2/3 + 2\nu(n)/3})$ queries.
\end{lemma}

\begin{proof}
    Consider a deterministic non-adaptive algorithm which queries $k=o(n^{2/3+2\nu(n)/3})$ entries, and let $E \subset [n] \times [n]$ be this set of queries---up to a factor of $2$, we may assume that the algorithm is symmetric so that it queries $(i,j)$ and $(j, i)$. The probability that a  violating triangle is among the queries $E$, via a union bound, is at most
    \[ \sum_{\substack{a \in A \\ x \in X}} \Prx_{\bpi}\left[ \left\{ \begin{array}{c} (\bpi(a), \bpi(a+x)), \\ (\bpi(a+x), \bpi(a+2x)), \\ (\bpi(a), \bpi(a+2x)) \end{array} \right\} \subset E\right], \]
    where the summation is over the $n|X|$ possible violating triangles. If we let $T(E)$ denote the set of triangles among the $E$ queries, the above expression becomes
    \[ n|X| \cdot O\left(\dfrac{1}{n^3}\right) \cdot |T(E)|,\]
    since the probability that any fixed triangle is mapped to a fixed triangle in $T(E)$ under $\bpi$ is $O(1/n^3)$. Since the maximum number of triangles in $|E|$ edges is $O(|E|^{3/2})$, the probability is upper bounded by
    \[ O\left(\frac{|X|}{n^2} \cdot |E|^{3/2}\right) = O\left(\frac{|E|^{3/2}}{n^{1 + \nu(n)}}\right),\]
    which is $o(1)$ if $|E| = o(n^{2/3 + 2\nu(n)/3})$.

\end{proof}

\newcommand{\BG}{\textsc{BG}}


\ignore{In this section, we show how Theorem~3 in Section 6 of~\cite{AKKR08} reduces the problem of finding a two-sided error metric testing lower bound to the problem of one-sided error testing lower bound. Since the above section incorporates a setting of weights to their construction graph construction (the graphs $\BG(n,|X|)$ from Section 5.3 in~\cite{AKKR08}), the argument below simply adapts the reduction to our setting with additional weights. 

\begin{lemma}\label{lem:algo-equiv}
    Consider the distribution $\Dno$ on the $(3n)\times (3n)$ matrices (above) and the distribution $\BG(n,|X|)$ on graphs with $3n$ vertices in~\cite{AKKR08}. We have:
    \begin{enumerate}
        \item Suppose there exists a non-adaptive algorithm that makes $q(n)$ queries and finds a violating triangle in a draw $\bM\sim \Dno$ with success probability at least 2/3. Then there exists a non-adaptive algorithm which makes $O(q(n))$ queries and finds a triangle in $\bG\sim \BG(n,|X|)$ with success probability at least 2/3.
        \item Suppose there exists a non-adaptive algorithm that makes $q(n)$ queries and finds a triangle in a draw $\bG\sim \BG(n,|X|)$ with success probability at least 2/3. Then there exists a non-adaptive algorithm that makes $O(q(n))$ queries and finds a violating triangle in $\bM\sim\Dno$ with success probability at least 2/3.
    \end{enumerate}
    The metric testing algorithm has query access to entries of a matrix $\bM\sim\Dno$; the triangle-free testing algorithm has query access to entries of the adjacency matrix of $\bG\sim \BG(n,|X|)$.
\end{lemma}
\begin{proof}
    By constructions of $\BG(n,|X|)$ and $\Dno$, there is a natural bijective map between matrices in $\Dno$ and graphs in $\BG(n,|X|)$ such that three indices $\{i,j,k\}\in[n]$ form a violating triangle in a matrix $\bM$ from $\Dno$ if and only if the three indices form a triangle in the corresponding graph in $\BG(n,|X|)$. 
    
    Suppose $\calA^M$ is the metric tester proposed in the first item. Suppose $A(\bG)$ is the adjacency matrix of the graph $\bG$ drawn from $\BG(n,|X|)$. Define a tester $\calA^G$ such that whenever $\calA^M$ queries an entry $\bM(i,j)$, $\calA^M$ queries the entry $A(\bG)(i,j)$. Since $\calA^M$ is non-adaptive, the queries do not depend on the answers of the previous queries. With probability at least 2/3, over the draw $\bM\sim \Dno$, $\calA^M$ detects some violating triangle in the matrix $\bM$, and so $\calA^G$ detects a triangle in the graph $G$ corresponding to $\bM$. Since both distributions $\Dno$ and $\BG(n,|X|)$ sample matrices and graphs from their support uniformly at random, $\calA^G$ detects a triangle in the draw $\bG\sim \BG(n,|X|)$ with probability at least 2/3.

    The second item follows a similar argument. For the proposed triangle-free tester $\calA^G$, define a tester $\calA^M$ such that, whenever $\calA^G$ queries the entry $A(\bG)(i,j),$ $\calA^M$ queries the entry $\bM(i,j)$. Thus, $\calA^M$ detects a violating triangle with probability at least 2/3 over the draw of $\bM\sim\Dno$.
\end{proof}

\begin{lemma}\label{lem:lb-2}
    Let $\Dno$ be the distribution on $(3n)\times(3n)$ as defined before. Over the draw of $\bM\sim \Dno$, any two-sided error non-adaptive metric testing algorithm with success probability at least 5/6 must make $\Omega(n^{2/3+2\nu(n)/3})$ queries.
\end{lemma}
\begin{lemma}[Theorem~4 in~\cite{AKKR08}]\label{lem:one-to-two}
    Let $\calD_\Delta$ be a distribution over graphs with $n$ vertices, and let $q(n)$ be a function of $n$. Assume the following holds:
    \begin{enumerate}
        \item With probability $1-o(1)$ a graph selected according to $\calD_\Delta$ is $\eps$-far from being triangle-free for some $\eps$.
    \item In all graphs in the support of $\calD_\Delta$, the triangles are edge-disjoint, and for any algorithm $\calA$, the probability that $\calA$ reveals a triangle in a graph selected according to $\calD_\Delta$ using $o(q(n))$ queries is less than 2/3.
    \end{enumerate}
Then any two-sided error algorithm for testing triangle-freeness that has success probability at least 5/6 must
perform $\Omega(q(n/2))$ queries.
\end{lemma}
\begin{proof}[Proof of Lemma~\ref{lem:one-to-two}]
    The proof is in~\cite{AKKR08}.
\end{proof}
\begin{proof}[Proof of Lemma~\ref{lem:lb-2}]
    It was shown in Lemma~\ref{lem:algo-lb-1} that, over the draw of $\bM\sim \Dno$, any non-adaptive algorithm $\calA^M$ that makes $o(n^{2/3+2\nu(n)/3})$ queries to $\bM$ finds a violating triangle with probability less than 2/3. By Lemma~\ref{lem:algo-equiv}, any non-adaptive algorithm $\calA^G$ that makes $o(n^{2/3+2\nu(n)/3})$ queries to $\bG\sim \BG(n,|X|)$ finds a triangle with probability less than 2/3. Moreover, with probability 1, $\bG\sim \BG(n,|X|)$ is $|X|/n$-far from being triangle-free, as constructed in~\cite{AKKR08}. Following Lemma~\ref{lem:one-to-two}, any two-sided error non-adaptive algorithm for testing triangle-freeness of graphs in $\BG(n,|X|)$ that has success probability at least 5/6 must perform $\Omega(n^{2/3+2\nu(n)/3})$ queries. Again using Lemma~\ref{lem:algo-equiv}, any two-sided error non-adaptive metric testing algorithm with success probability at least 5/6 must make $\Omega(n^{2/3+2\nu(n)/3})$ queries.
\end{proof}}

\newcommand{\UltraTesting}{\textsc{UltraTesting}}
\newcommand{\TreeTesting}{\textsc{TreeTesting}}
\newcommand{\x}{x}

\section{Ultrametric and Tree Metric Testing Upper Bound: Theorem~\ref{thm:testing-trees}} \label{sec:ultra}

In this section, we present algorithms for testing ultrametrics and tree metrics, both using $\tilde{O}(1/\eps)$ samples and $\tilde{O}(1/\eps^2)$ queries, thereby proving Theorem~\ref{thm:testing-trees}. We let
\begin{align*}
\calP^U &=\{M\in\calC,M\text{ encodes an ultrametric space over }[n]\}, \\
\calP^T &=\{M\in\calC,M\text{ encodes a tree metric space over }[n]\}.
\end{align*}
\subsection{Ultrametric Testing Upper Bound}
As per Lemma~\ref{lem:clean}, input matrices $M\in\calC$ are already symmetric, non-negative, and zero only on the diagonal. As in Section~\ref{sec:general}, algorithms will have one-sided error, meaning that the tester's task is finding a certificate that the input matrix is not in $\calP^{U}$. We first define the type of violation the algorithm seeks.
\begin{Definition}[Ultrametric Violating Triple]\label{def:vio-ultra}
    Given $M\in \calC$, the triple $\{i,j,k\}$ is a violation of ultrametric if, after renaming so $M(i,j)$ is maximum among the 3 pairwise distances,
    \begin{align*}
         M(i,j)>M(i,k)\text{ and }M(i,j)>M(j,k).
    \end{align*}
\end{Definition}

We present the algorithm for testing ultrametrics, $\UltraTesting$. 
\begin{framed}
    \textbf{$\UltraTesting$ Algorithm}. The algorithm aims to find a violating triple of indices $\{i,j,k\}$ of the ultrametric property. If a violating triple is found, it constitutes a certificate that the matrix is not ultrametric, and the algorithm outputs ``reject".

    \textbf{Input:} The parameters $n\in\mathbb{N}$ and $\eps\in(0,1)$, as well as query access to the entries of an unknown
    $n \times n$ matrix $M$ from $\calC$ (see Lemma~\ref{lem:clean}).

    \textbf{Output:} ``accept" or ``reject."

    \begin{enumerate}
        \item\label{en:step1} For $s = \Theta(\log(1/\eps)/\eps)$, take $s$ random samples $\bi_1,\dots, \bi_s \sim [n]$ drawn independently. Query $M(\bi_{\ell}, \bi_{k})$ for all $\ell, k \in [s]$. 
        \item\label{en:step2} If there exists a triple among the sampled indices $\{ \bi_{\ell}, \bi_{k}, \bi_{h} \}$ which is a violating triple in $M$, output ``reject.''
    \end{enumerate}
\end{framed}

\begin{lemma}[Ultrametric Testing Algorithm Lemma]\label{lem:ultra-algo}
    For $n \in \N$ and $\eps \in (0,1)$, the algorithm $\UltraTesting$ receives as input an $n\times n$ matrix $M \in \calC$ and a parameter $\eps$ and has the following guarantees:
\begin{itemize}
\item If $M \in \calP^U$, $\UltraTesting(M,\eps)$ always outputs ``accept.''
\item If $M\in\calC$ is $\eps$-far from $\calP^U$, the algorithm outputs ``reject'' with probability at least $2/3$.
\end{itemize}
The algorithm is non-adaptive, taking $O(\log(1/\eps)/\eps)$ samples and using $O(\log(1/\eps)^2/\eps^2)$ queries.
\end{lemma}


\subsubsection{Ultrametric Testing Algorithm: Proof of Lemma~\ref{lem:ultra-algo}}

For the sake of analysis, we will divide the samples obtained in Line~\ref{en:step1} of $\UltraTesting$ into two distinct groups. 
\begin{itemize}
\item The first group consists of a set $\bS = \{ \bi_1,\dots, \bi_{s/2} \}$ of the first $O(\log(1/\eps)/\eps)$ samples. As in~\cite{PR01}, we use the set $\bS$ to define a ``skeleton partition'' (Definition~3.5 in~\cite{PR01} and Definition~\ref{def:skeleton-separator} below).
\item The second group consists of the remaining $O(\log(1/\eps) / \eps)$ samples viewed as a set of consecutive pairs, since the violations we find are formed by such some pairs $(\bi_{\ell}, \bi_{\ell+1})$ violating a constraint imposed by the skeleton partition of $\bS$ in the first group.
\end{itemize}
We begin by defining the skeleton partition and corresponding equivalence classes formed by a fixed subset $S\in[n]$. Our analysis will track how the equivalence classes evolve as the (random) set changes with each sample. 
\newcommand{\SEP}{\textsc{SEP}}
\newcommand{\SC}{\textsc{SC}}
\newcommand{\EC}{\textsc{EC}}
\begin{Definition}[Skeleton Partition and Equivalence Classes]\label{def:skeleton-separator}
    Let $M \in \calC$ and $S \subset [n]$. We say that $S$ is consistent if the $|S| \times |S|$ submatrix $M_{|S \times S}$ encodes an ultrametric. For a consistent set $S$, we define:
    \begin{itemize}
        \item \emph{\textbf{Consistent Points}}. A point $j \in [n] \setminus S$ is a consistent point if $S \cup \{j\}$ is consistent. 
        \item \emph{\textbf{Skeleton Partition}}. A skeleton partition of $S$ is a partition $P_1,\dots, P_{\ell}$ of the consistent points in $[n] \setminus S$ where two indices $j,k\in[n]\setminus S$ are in the same part $P_{p}$, or equivalence class, if and only if $M(j, i)=M(k, i)$ for all $i\in S$.
        \item \emph{\textbf{Separator Set}}. If $j,k \in [n] \setminus S$ are in different equivalence classes, their separator set $\SEP(j,k) \subset S$ is given by
        \[ \SEP(j,k) = \left\{ i \in S : M(i,j) \neq M(i,k) \right\}. \]
        \item \emph{\textbf{Separator Corruption}}. If $j,k \in[n]\setminus S$ are in different equivalence classes, the pair $(j,k)$ is a separator corruption if there exists $i \in \SEP(j,k)$ where 
        \begin{align}
            M(j,k) &\neq \max\left\{ M(i,j), M(i,k) \right\}. \label{eq:sep-vio}
        \end{align}
        We let $\SC(M, S) \subset ([n] \setminus S) \times ([n] \setminus S)$ denote the set of pairs which are separator corruptions.
    \end{itemize} 
\end{Definition}

Definition~\ref{def:skeleton-separator} suggests our analysis approach. The algorithm will sample a set $\bS$ of indices for the first group---if the set $\bS$ is not consistent, there is a violation and we are already done. Assume that $\bS$ is consistent. We aim to sample a pair $(\bj, \bk)$ in the second group forming a separator corruption. Note that (\ref{eq:sep-vio}) forms a violation of the ultrametric property, so if this occurs, the algorithm can safely output ``reject.''
We define one more type of ``easy-to-detect'' violation to the ultrametric property. 

\begin{Definition}[Easy-to-Detect Corruption]
    For $M \in \calC$ and a consistent set $S\subset[n]$, let $(j,k) \in ([n] \setminus S) \times ([n] \setminus S)$ be in the same equivalence class. The pair $(j,k)$ form an easy-to-detect corruption if there exists $i \in S$ where
    \[ M(j,k) > M(j,i) = M(k,i).\]
    Let $\EC(M,S) \subset ([n] \setminus S) \times ([n] \setminus S)$ be set of easy-to-detect corruptions in $M$ with respect to $S$.
\end{Definition}

From the analysis perspective, we prove the following lemma, which implies the correctness of our algorithm. We state Lemma~\ref{lem:corruptions}, as it will directly imply Lemma~\ref{lem:ultra-algo}. The proof is a straightforward consequence of two lemmas (Lemma~\ref{lem:large-EC-ODC} and Lemma~\ref{lem:AS-large}), which encapsulates our improvement over the analysis of~\cite{PR01}.
\begin{lemma}\label{lem:corruptions}
    Let $M \in \calC$ be $\eps$-far from $\calP^U$. Then, with probability at least $5/6$ over the draw of $\bS = \{ \bi_1,\dots, \bi_{s/2} \} \subset [n]$ where $\bi_1,\dots, \bi_{s/2} \sim [n]$ and $s = O(\log(1/\eps)/\eps)$, one of the followings holds:
    \begin{itemize}
        \item $\bS$ is not consistent;
        \item There are at least $\eps n/32$ points in $[n] \setminus \bS$ which are not consistent with $S$; 
    \item $|\SC(M, \bS)| + |\EC(M,\bS)| \geq \eps n^2/8. 
    $
    \end{itemize}
\end{lemma}

\begin{proof}[Proof of Lemma~\ref{lem:ultra-algo} assuming Lemma~\ref{lem:corruptions}] 
First, if $M\in\calP^U$, $\UltraTesting$ never observe a violation and outputs ``accept.'' If $M \in \calC$ is $\eps$-far from $\calP^U$, we use Lemma~\ref{lem:corruptions} to deduce that with probability at least $5/6$, the set $\bS = \{ \bi_1,\dots, \bi_{s/2}\}$ obtained from the first $s/2$ samples in Line~\ref{en:step1} is either already inconsistent (in which case among the entries of $\bS$, there already is a violation), or it will become easy to sample a violation. 
Assume that $\bS$ is consistent and that either there are $\eps n / 32$ inconsistent points in $[n] \setminus \bS$, or $|\SC(M,\bS)| + |\EC(M,\bS)| \geq \eps n^2/8$. We turn to the second group of $s/2$ samples, which are divided into pairs $(\bj, \bk)$. We note that in order for a violation to be avoided, the $\Theta(\log(1/\eps)/\eps)$ pairs of samples must all avoid the $\eps n/32$ inconsistent points, and, at the same time, avoid the $\eps n^2 / 8$ entries in $\SC(M,\bS)$ and $\EC(M,\bS)$. The probability this occurs is at most   
\[ \left(1 - \Omega(\eps) \right)^{\Theta(\log(1/\eps)/\eps)} + \left(1 - \Omega(\eps) \right)^{\Theta(\log(1/\eps)/\eps)} \leq 1/6. \]

for an appropriate choice of the constant in front of the sample size.

Thus, by union bound over all events, the probability that the algorithm does not find any violation will be at most $1/3$, which completes the proof.
\end{proof}

We turn our attention to proving Lemma~\ref{lem:corruptions}, which will follow by lower-bounding the sizes of the separator corruptions $\SC(M,\bS)$ and easy-to-detect corruptions $\EC(M,\bS)$. Instead of analyzing $\SC(M,\bS)$ and $\EC(M,\bS)$ directly, we will look at a related quantity defined over certain types of equivalence classes.

\begin{Definition}[Easy, Versatile, and Active Parts]\label{def:parts}
    Let $M\in\calC$ be $\eps$-far from $\calP^U$ and $S \subset [n]$ be consistent, and let $P_1,\dots, P_{\ell}$ denote the skeleton partition. 
    \begin{itemize}
        \item \textbf{\emph{Easy Part}}. A part $P_{\ell}$ is \emph{easy} if 
        \[ \Prx_{\bj, \bk \sim P_{\ell}}\left[(\bj,\bk) \in \EC(M, \bS) \right] \geq \frac{1}{2}.  \]
        That is, at least half of the entries in the submatrix $M_{|P_\ell\times P_\ell}$ are easy-to-detect corruptions.
        \item \textbf{\emph{Versatile Part}}. Consider a part $P_{\ell}$ with $\beta n$ indices 
        for $\beta = \Omega (1)$. 
        We say $P_{\ell}$ is \emph{versatile} if the following hold:
        \begin{itemize}
            \item there exists a $(\beta n)\times(\beta n)$ matrix $M^\ast_{P_\ell}$ encoding an ultrametric whose entries are at most $\min_{i\in S,j\in P_l}\{M(i,j)\}$.
            \item $\|M^\ast_{P_\ell}-M_{|P_{\ell} \times P_{\ell}}\|\leq \beta\cdot \eps n^2/2$.
        \end{itemize}
        That is, intuitively, no more than $\beta\cdot \eps n^2/2$ entries of the submatrix $M_{|P_{\ell} \times P_{\ell}}$ need to be modified so that $M_{|P_{\ell} \times P_{\ell}}$ is fixed to be an ultrametric consistent with $S$.
        \item \textbf{\emph{Active Part}}. A part $P_{\ell}$ is \emph{active} if it is neither easy nor versatile. We let \
        \begin{align*} 
        A(M, S) &= \left\{ (j,k) \in P_{\ell} \text{ for some active part $P_{\ell}$} \right\} \\
        \alpha(M, S) &= \Prx_{\bj \sim [n] }\left[ \bj \in P_{\ell} \text{ for some active part $P_{\ell}$}\right]
        \end{align*}
    \end{itemize}
\end{Definition}

\begin{lemma}\label{lem:large-EC-ODC}
    Let $M\in\calC$ be $\eps$-far from $\calP^U$ and $S \subset [n]$ be consistent. If $|A(M, S)|\leq \eps n^2/8$, then either there are at least $\eps n/32$ inconsistent points in $[n]\setminus S$, or $|\SC(M, S)| + |\EC(M,S)| \geq \eps n^2/8$.
\end{lemma}

\begin{proof}
    We prove the contra-positive by showing that, given $M$ being $\eps$-far from $\calP^U$ and a consistent subset $S \subset [n]$ which satisfy (i) $|A(M, S)| \leq \eps n^2/8$, (ii) $|\SC(M,S)| + |\EC(M,S)| \leq \eps n^2/8$, and (iii) at most $\eps n / 32$ inconsistent points in $[n] \setminus S$, there exists an $n \times n$ matrix $\tilde{M} \in \calP^{U}$ which differs from $M$ on fewer than $\eps n^2$ entries. This implies that $M$ is $\eps$-close to $\calP^U$, which is a contradiction. The matrix $\tilde{M}$ is constructed as follows:
    \begin{enumerate}
        \item For all $i, i' \in S$ we set $\tilde{M}(i,i')$ to $M(i,i')$.
        \item\label{en:second-change} If $j \in [n] \setminus S$ is among the $\eps n /32$ inconsistent points, we let $\tilde{M}(j,k)$ be some arbitrarily large value for all $k\in[n]$. This corresponds to effectively removing $j$ while keeping an ultrametric. For the points $j \in [n] \setminus S$ which are consistent, set $\tilde{M}(j,i)$ to be $M(j,i)$ for all $i \in S$.
        \item\label{en:third-change} For consistent $j,k \in [n] \setminus S$ lying in different equivalence classes:
        \begin{itemize}
            \item If $(j,k) \notin \SC(M,S)$, set $\tilde{M}(j,k)=M(j,k)$. 
            \item If $(j,k)\in \SC(M,S),$ find $i\in\SEP(j,k)$ and set $\tilde{M}(j,k)=\max\{M(i,j),M(i,k)\}$. (As we will see, the specific choice of $i$ will not matter).
        \end{itemize}
        \item For consistent $j,k\in[n]\setminus S$ lying in the same equivalence class $P_{\ell}$:
        \begin{itemize}
            \item If $P_{\ell}$ is an Easy Part or an Active Part, set $\tilde{M}(j,k)$ to be the minimum positive entry in $M$.
            \item If $P_{\ell}$ is a Versatile Part, find the matrix $M^\ast_{P_{\ell}}$ as in Definition~\ref{def:parts}, and set $\tilde{M}(j,k)=M^\ast_{P_{\ell}}(j,k)$.
        \end{itemize}

    \end{enumerate}

    We now show that $\|\tilde{M}-M\|_{0}<\eps n^2$ and $\tilde{M}\in \calP^U$. First, note that $M$ and $\tilde{M}$ differ on at most
    \[ \frac{\eps n}{32} \times 2n + |\SC(M,S)| + \sum_{\substack{P_{\ell} \text{ Easy}}} |P_{\ell}|^2 + |A(M,S)| + \sum_{P_{\ell} \text{ Versatile}} |P_{\ell}| \cdot \eps n / 2 \] entries.
    Notice that, $2|\EC(M,S)|$ upper bounds the sum of $|P_{\ell}|^2$ over Easy Part $P_{\ell}$ (since each Easy Part $P_{\ell}$ has at least half of its entries in $\EC(M,S)$. Furthermore, parts partition consistent points in $[n] \setminus \bS$, so the sum of $|P_{\ell}|$ over Versatile parts is at most $n$. Finally, the assumption $|\SC(M, S)| + |\EC(M,S)| \leq \eps n^2 / 8$ as well as $|A(M,S)| \leq \eps n^2 / 8$ give the desired bound: $\|\tilde{M}-M\|_0\leq 15/16\cdot\eps n^2$.

    The rest of the proof which shows that $\tilde{M}\in\calP^U$ has been moved to the appendix, in the interest of space. We prove $\tilde{M}\in\calP^U$ by considering an arbitrary triple $\{i, j, k\} \subset [n]$ and showing that it does not form a violation in $\tilde{M}$.

    The above argument shows that $\tilde{M}$ encodes an ultrametric and is $(15/16\cdot\eps)$-close to $M$, which leads to a contradiction. Hence, if $|A(M,S)|\leq \eps n^2/8$, then either there are at least $\eps n^2/32$ inconsistent points in $[n]\setminus S$ or $|\SC(M,S)|+|\EC(M,S)|\geq \eps n^2/8$.
    
\end{proof}

\begin{lemma}\label{lem:AS-large}
    Suppose $M\in\calC$ is $\eps$-far from $\calP^U$. With probability at least $5/6$, over the draw of $\bS\subset[n]$ of size $\Omega(\log(1/\eps)/\eps)$, $|A(M,\bS)|\leq \eps n^2/8$.
\end{lemma}

\begin{proof}[Proof of Lemma~\ref{lem:corruptions} assuming Lemmas~\ref{lem:AS-large}]
The proof is a straightforward application of Lemma~\ref{lem:AS-large} and Lemma~\ref{lem:large-EC-ODC}. In particular, by Lemma~\ref{lem:AS-large}, a draw of $\bS$ will satisfy $|A(M,\bS)| \leq \eps n^2 / 8$ with probability at least $5/6$. Then, either $\bS$ is inconsistent, or we may apply Lemma~\ref{lem:large-EC-ODC}, which implies either there are at least $\eps n/32$ indices which incur violations with $S$ or $|\SC(M,\bS)| + |\EC(M,\bS)| \geq \eps n^2 / 8$.
\end{proof}

\subsubsection{Bounding the Number of Active Entries: Proof of Lemma~\ref{lem:AS-large}}

\begin{lemma}\label{lem:active-entry-reduce_AS}
    Suppose $M\in\calC$ is $\eps$-far from $\calP^U$ and suppose $S=\{i_1,...,i_{j-1}\}\subset[n]$. Over the randomness of the index $\bi_j$ sampled from $[n]$, $\Ex_{\bi_j}[|A(M,S\cup\{\bi_{j}\})|]\leq |A(M,S)|\cdot\exp(-\frac{\eps}{16}).$
\end{lemma}
\begin{proof}[Proof of Lemma~\ref{lem:AS-large} assuming Lemma~\ref{lem:active-entry-reduce_AS}]
    
    Suppose $\bi_1,...,\bi_s\in[n]$ are drawn independently where $s=16\log(48/\eps)/\eps$. Let $\bS$ be this set of random samples. Then the law of total expectation, 
    \begin{align*}
        \log\left(\Ex_{\bi_1,...,\bi_s}[|A(M,\bS)|]\right)=&\log\left(\Ex\left[|A(M,\{\bi_1,...,\bi_{s-1}\}\cup\{\bi_s\})|\right]\right)\\
        =&\log\left(\Ex\left[\Ex_{\bi_{s}}\left[|A(M,\{\bi_1,...,\bi_{s-1}\}\cup\{\bi_s\})||\bi_1,...,\bi_{s-1}\right]\right]\right)\\
        \leq&
        \log\left(\Ex\left[|A(M,\{\bi_1,...,\bi_{s-1}\})|\cdot\exp(-\frac{\eps}{16})\right]\right)\\
        =&
        \log\left(\Ex_{{\bi_1,...,\bi_{s-1}}}\left[|A(M,\{\bi_1,...,\bi_{s-1}\})|\right]\right)-\frac{\eps}{16}.
    \end{align*}
    By induction, 
    \[
    \log\left(\Ex_{\bi_1,...,\bi_s}[|A(M,\bS)|]\right)=\log(\Ex[|A(M,\emptyset)|])-\frac{\eps\cdot s}{16}=\log(n^2)-\log(48/\eps)=\log(\eps n^2/48).
    \]
    Thus, $\Ex_{\bi_1,...,\bi_s}[|A(M,\bS)|]=\eps n^2/48.$ By Markov's inequality, with probability at least $5/6$, over the draw of $\{\bi_1,...,\bi_s\},$ $A(|M,\bS|)\leq \eps n^2/8$.


\end{proof}
\begin{lemma}\label{lem:sample-active}
    Suppose $M\in\calC$ is $\eps$-far from $\calP^U$. Suppose $S=\{i_1,...,i_{j-1}\}\subset[n]$. Over the randomness of the next sampled index $\bi_j$, if $\bi_j$ is sampled from an Active Part $P$ with size $\beta n$, 
    \[\Ex_{\bi_j}[|A(M,S\cup\{\bi_j\})||\bi_j\in P]\leq |A(M,S)|-\beta\cdot \eps n^2/16.
    \]
\end{lemma}
\begin{proof}[Proof of Lemma~\ref{lem:active-entry-reduce_AS} assuming Lemma~\ref{lem:sample-active}]

    Let $\calA(M,S)$ be the set of all Active Parts defined over the set $S$. Suppose the part $P_l$ with $\beta_l n$ points is active. A point is chosen from this part with probability $\beta_l$, and as a result, $\Ex_{\bi_j}[|A(M,S\cup\{\bi_j\})||\bi_j\in P_l]\leq |A(M,S)|-\beta_l\cdot \eps n^2/16$. Thus, we have the conditional expectation  
    \begin{align*}
    \Ex_{\bi_j}[|A(M,S\cup\{\bi_j\})||\bi_j\in \calA(M,S)]=&\sum_{l:P_l\in\calA(M,S)}\Ex_{\bi_j}[|A(M,S\cup\{\bi_j\})||\bi_j\in P_l]\cdot\Pr(\bi_j\in P_l|\bi_j\in\calA(M,S))
    \\\leq&\,\,
    |A(M,S)|-\sum_{l:P_l\in\calA(M,S)}\beta_l\cdot \epsilon n^2/16\cdot\frac{\beta_l}{\sum_{k:P_k\in\calA(M,S)}\beta_k}
    \\=&\,\,
    |A(M,S)|-\frac{\sum_{l:P_l\in\calA(M,S)}\beta_l^2\cdot \epsilon n^2/16}{\sum_{l:P_l\in\calA(M,S)}\beta_l}
    \end{align*}
    On the other hand, we have
    \[
    |A(M,S)|=\sum_{l:P_l\in\calA(M,S)}(\beta_ln)^2,
         \]
         \[
         \alpha(M,S)= \Prx_{\bj \sim [n] }\left[ \bj \in P_{\ell} \text{ for some active part $P_{\ell}$}\right]=\sum_{l:P_l\in\calA(M,S)}\beta_l.
    \]
    Hence, as index $\bi_j$ is sampled from some Active Part, $\Ex_{\bi_j}[|A(M,S\cup\{\bi_j\})||\bi_j\in \calA(M,S)]\leq |A(M,S)|-|A(M,S)|\cdot\frac{\eps}{16\cdot\alpha(M,S)}.$
    In general when index $\bi_j$ is sampled uniformly at random,
    \begin{align*}
        &\Ex_{\bi_j}\left[|A(M,S\cup\{\bi_j\})|\right]\\
        =&\Ex_{\bi_j}\left[|A(M,S\cup\{\bi_j\})||\bi_j\in \calA(M,S)\right]\cdot\Pr(\bi_j\in\calA(M,S))
        \\&+\Ex_{\bi_j}[|A(M,S\cup\{\bi_j\})||\bi_j\notin \calA(M,S)]\cdot\Pr(\bi_j\notin\calA(M,S))\\
        \leq&\left(|A(M,S)|-|A(M,S)|\cdot\frac{\eps}{16\cdot\alpha(M,S)}\right)\cdot \alpha(M,S)+|A(M,S)|\cdot(1-\alpha(M,S))\\
        =&|A(M,S)|\cdot\left(1-\frac{\eps}{16}\right)
        \leq |A(M,S)|\cdot\exp\left(-\frac{\eps}{16}\right)
    \end{align*}
\end{proof}
\begin{proof}[Proof of Lemma~\ref{lem:sample-active}]
    The Active Part $P$ corresponds to a sub-square-matrix $B=M_{|P\times P}$ of $M$. Consider all the entries in $B$ that are not in $\EC(M,S)$. Suppose the set of values these entries take on is $\{v_1,v_2,...,v_k\},$ and further suppose $v_1$ is the value that most entries take on. Suppose $\{v_{k+1},...,v_{l}\}$ is the set of values the entries in $\EC(M,S)\cap B$ take on. Following the definition of easy-to-detect corruption $\EC(M,S)$, any value in $\{v_{k+1},...,v_{l}\}$ is larger than any value in $\{v_1,v_2,...,v_k\},$ so $\{v_1,...,v_l\}$ is a set of non-repetitive values. Let $r_\tau$ to denote the number of entries taking on value $v_\tau$ in the block matrix $B$. Hence, $\sum_{\tau=1}^lr_\tau=(\beta n)^2.$ The assumption indicates that $r_1\geq r_\tau,\forall \tau\in\{2,...,k\},$ and $\sum_{\tau=k+1}^lr_\tau$ is the number of easy-to-detect-corruptions in $B$. That is,
    \[
    |B\cap \EC(M,S)|=\sum_{\tau=k+1}^lr_\tau.
    \]
    Since $P$ is an Active Part, it is not an Easy Part. By definition, no more than half of the entries in $B$ are in $\EC(M,S)$. Hence, $\sum_{\tau=k+1}^lr_\tau$ is no larger than $(\beta n)^2/2$.

Now, focus on the row/column in $B$ that corresponds to index $\bi_j$ in $P$. Suppose in the row, there are $r_l^{(\bi_j)}$ entries whose values are $v_l$. Then, once $\bi_j$ is sampled, some pairs of indices are separated into different parts. In particular, for indices $x,y\in P$, if $M(\bi_j,x)\neq M(\bi_j,y)$, the pair $x,y$ that previously both belonged to part $P$ now fall into two different parts. The number of such pairs of indices that are separated is $\sum_{\tau=1}^lr_\tau^{(\bi_j)}(\beta n-r_\tau^{(\bi_j)})/2.$ Over the randomness of sampling any index $\bi$ in $P$, the expected number of separated pairs created is 
\[
\frac{1}{\beta n}\sum_{\bi\in P}\sum_{\tau=1}^lr_\tau^{(\bi)}(\beta n-r_\tau^{(\bi)})/2=\frac{1}{2\beta n}\sum_{\tau=1}^l\sum_{\bi\in P}r_\tau^{(\bi)}(\beta n-r_\tau^{(\bi)}).
\]

Notice here that the number of separated pairs is exactly the number of entries moving from the block matrix $B$ to off-diagonal, and these entries no longer belong to Active Parts. Therefore, $|A(M,S)|-|A(M,S\cup\{\bi\})|$ is at least the number of separated pairs created by sampling an index $\bi\in P$. 

Hence, the goal is to find a lower bound for the number of separated pairs created by an index $\bi\in P$. Since a pair of indices are separated if their distance value to the sampled index $\bi$ is different, we focus on the values in the matrix. Imagine we have a graph with $\beta n$ vertices, and two vertices $\bi_x,\bi_y$ are connected by an edge if and only if the entry $(\bi_x,\bi_y)$ in $B$ equals $v_\tau$. Then there are $r_\tau/2$ edges in this graph, and the degree of a vertex $\bi$ is $r_\tau^{(\bi)}$. Thus, 
\[
\sum_{\bi\in P}r_\tau^{(\bi)}(\beta n-r_\tau^{(\bi)})=\beta n\cdot\sum_{\bi\in P}r_\tau^{(\bi)}-\sum_{\bi\in P}(r_\tau^{(\bi)})^2=\beta n\cdot r_\tau-\sum_{\bi\in P}(r_\tau^{(\bi)})^2.
\]
De Caen~\cite{D} gives an upper bound on the sum of squared degrees, that 
\[
\sum_{\bi\in P}(r_\tau^{(\bi)})^2\leq E({2E}/({V-1})+V-2)\approx{r_\tau^2}/{(2\beta n)}+r_\tau\cdot\beta n/2.
\]
On the other hand, since $r_1\geq r_\tau,\forall \tau\in\{2,...,k\},$ an upper bound for $\sum_{\tau=1}^lr_\tau^2$ is 
\[
\sum_{\tau=1}^lr_\tau^2
\leq \left(\sum_{\tau=k+1}^lr_\tau\right)^2+\left((\beta n)^2-\sum_{\tau=k+1}^lr_\tau\right)r_1,
\]
where, recall that, the quantity $\sum_{\tau=k+1}^lr_\tau$ is the total number of easy-to-detect corruptions in $B$. To sum up, $|A(M,S)|-|A(M,S\cup\{\bi\})|$ is at least the number of pairs being separated. On the other hand, in expectation over the randomness of the sampled index $\bi\in P$, the number of pairs being separated is at least 
\begin{align*}
    &\frac{1}{2\beta n}\sum_{\tau=1}^l\sum_{\bi\in P}r_\tau^{(\bi)}(\beta n-r_\tau^{(\bi)})
    \\\geq&\frac{1}{2\beta n}\sum_{\tau=1}^lr_\tau\cdot\beta n-\frac{r_\tau^2}{2\beta n}-r_\tau\cdot\beta n/2
    \\=&\frac{1}{2\beta n}(\beta n)^3-\frac{1}{4(\beta n)^2}\sum_{\tau=1}^lr_\tau^2-\frac{(\beta n)^2}{4}
    \\\geq&\frac{(\beta n)^2}{4}-\frac{1}{4(\beta n)^2}\left(\left(\sum_{\tau=k+1}^lr_\tau\right)^2-r_1\cdot\sum_{\tau=k+1}^lr_\tau\right)-\frac{r_1}{4}
\end{align*}

To further bound this quantity, we consider two cases for the size of $r_1$, i.e. the number of entries taking on value $v_1$.
\begin{itemize}
    \item CASE $r_1\leq (\beta n)^2/2:$ Following the assumption that $P$ is not an Easy Part, $\sum_{\tau=k+1}^lr_\tau$, which equals $|\EC(M,S)\cap B|$, shall not exceed $|B|/2=(\beta n)^2/2$. Thus, the above expected number of separated pairs is at least $(\beta n)^2/4-(\beta n)^4/(16(\beta n)^2)-r_1/4=3(\beta n)^2/16-r_1/4$. As a consequence, 
    \[
    8\cdot\left(|A(M,S)|-\Ex_{\bi}[|A(M,S\cup\{\bi\})||\bi\in P]\right)\geq3(\beta n)^2/2-2r_1\geq (\beta n)^2-r_1.
    \]
    \item CASE $r_1>(\beta n)^2/2$: Notice that $\sum_{\tau=k+1}^lr_\tau\leq (\beta n)^2-r_1<(\beta n)^2/2$. In this case, the above expected number of separated pairs is minimized when $\sum_{\tau=k+1}^lr_\tau$ is 0 or $(\beta n)^2-r_1$ by convexity. As a result, $|A(M,S)|-\Ex_{\bi}[|A(M,S\cup\{\bi\})||\bi\in P]\geq\max\{(\beta n)^2/4-r_1/4,r_1/2\}$. Again, it implies that 
    \[8\cdot\left(|A(M,S)|-\Ex_{\bi}[|A(M,S\cup\{\bi\})||\bi\in P]\right)\geq(\beta n)^2-r_1.\]
\end{itemize}

Now we show by contradiction that, since $P$ is not a Versatile Part, $\left(\sum_{\tau=2}^lr_\tau\right)=(\beta n)^2-r_1$ must be at least $\beta \cdot \eps n^2/2$. 
\begin{claim}
    Since $P$ is an Active Part, $\sum_{\tau=2}^lr_\tau=(\beta n)^2-r_1\geq \beta \cdot \eps n^2/2$.
\end{claim}
\begin{proof}
    Suppose $\sum_{\tau=2}^lr_\tau=(\beta n)^2-r_1 < \beta \cdot \eps n^2/2$. Create a $(\beta n)\times(\beta n)$ matrix $M^\ast_{P}$ where diagonal entries are 0 and every other entry is set to $v_1$. Then, $\|M^\ast_P-B\|\leq \sum_{\tau=2}^lr_\tau<\beta \cdot n^2/2$. Moreover, $M^\ast_P$ encodes an ultrametric. Lastly, as the value $v_1$ is taken on by some entry $M(u,u')\notin \EC(M,S)$, for any index $i\in S$, $M(i,u)=M(i,u')\geq M(u,u')=v_1$. Moreover, for any $u''\in P$, $M(i,u)=M(i,u'').$ Thus, $v_1\leq \min_{i\in S,u\in P}\{M(i,u)\}$. Therefore, $M^\ast_P$ satisfies all conditions that make $P$ a Versatile Part. However, because $P$ is an Active Part, by contradiction, we must have $\sum_{\tau=2}^lr_\tau=(\beta n)^2-r_1\geq \beta \cdot \eps n^2/2$.
\end{proof}


As a summary for the above arguments, we have 
\[
8\cdot\left(|A(M,S)|-\Ex_{\bi}[|A(M,S\cup\{\bi\})||\bi\in P]\right)\geq (\beta n)^2-r_1\geq \beta\cdot\epsilon n^2/2.
\]
\end{proof}

\newcommand{\tM}{\tilde{M}}

\subsection{Tree Metric Testing Upper Bound}\label{subsec:tree}

In this section, we finish the proof of Theorem~\ref{thm:testing-trees} by presenting a tree metric testing algorithm $\TreeTesting$. We first define the type of violation the algorithm seeks.

\begin{Definition}[Tree Metric Violating Quadruple]\label{def:vio-tree}
    For $i,j,k,l\in[n],$ the quadruple $\{i,j,k,l\}$ is a violation for tree metrics in $M\in \calC$ if, after re-naming so $M(i,j)+M(k,l)$ is the maximum among the three matchings
    \[
    \{ M(i,j)+M(k,l),M(i,k)+M(j,l), M(i,l)+M(j,k) \},
    \]
    \begin{align*}
         M(i,j)+M(k,l)>M(i,k)+M(j,l))\text{ and }M(i,j)+M(k,l)>M(i,l)+M(j,k).
    \end{align*}
\end{Definition}

The testing algorithm $\TreeTesting$ below is very similar to the ultrametric tester we presented earlier. The analysis follows an analogous path.

\begin{framed}
    \textbf{TreeTesting Algorithm}: The algorithm aims to find a violating quadruple of indices $\{i,j,k,l\}$ to the tree metric property Definition~\ref{def:vio-tree}. If such a quadruple is found, it constitutes a certificate that the matrix is not a tree metric, and the algorithm outputs ``reject".

    \textbf{Input:} The parameters $n\in\mathbb{N}$ and $\eps\in(0,1)$, as well as query access to the entries of an unknown
    $n \times n$ matrix $M$ from $\calC$ (see Lemma~\ref{lem:clean}).

    \textbf{Output:} ``accept" or ``reject"

    \begin{enumerate}
        \item\label{en:line1-tree} For $s = O(\log(1/\eps)/\eps)$, take $s$ random samples $\bi_1,\dots, \bi_s \sim [n]$ drawn independently. Query $M(\bi_{\ell}, \bi_{k})$ for all $\ell, k \in [s]$. 
        \item If there exists a violating quadruple among the sampled indices $\{ \bi_{a}, \bi_{b}, \bi_{c}, \bi_d \}$, output ``reject.''
        
    \end{enumerate}
\end{framed}

\begin{lemma}[Tree Metric Testing Algorithm Lemma]\label{lem:tree-algo}
    For $n \in \N$ and $\eps \in (0,1)$, there exists a randomized algorithm, $\TreeTesting$, which receives as input an $n\times n$ matrix $M \in \calC$ and a parameter $\eps$ and has the following guarantees:
\begin{itemize}
\item If $M \in \calP^T$, $\TreeTesting(M,\eps)$ always outputs ``accept.''
\item If $M\in\calC$ is $\eps$-far from $\calP^T$, the algorithm outputs ``reject'' with probability at least $2/3$.
\end{itemize}
The algorithm is non-adaptive, taking $O(\log(1/\eps)/\eps)$ samples and using $O(\log(1/\eps)^2/\eps^2)$ queries.
\end{lemma} 
\begin{proof}[Proof of Theorem~\ref{thm:testing-trees} with Lemma~\ref{lem:ultra-algo} and assuming Lemma~\ref{lem:tree-algo}]
By Lemma~\ref{lem:clean}, it suffices to show Theorem~\ref{thm:testing-trees} holds for $M\in\calC$.  
For $M\in\calP^T$ (respectively $M\in\calP^U$), $\TreeTesting(M,\eps)$ (resp. $\UltraTesting(M,\eps)$) outputs ``accept" with probability 1, so the algorithm outputs ``accept" if $M$ encodes a tree metric space (resp. ultrametric space). If $M$ is $\eps$-far from $\calP^T$ (resp. $\eps$-far from $\calP^U$), the $\TreeTesting(M,\eps)$ algorithm (resp. $\UltraTesting(M,\eps)$ algorithm) outputs ``reject" with probability at least 2/3. Both algorithms are non-adaptive with one-sided error and sample complexity $O(\log(1/\eps)/\eps)$ and query complexity $O(\log(1/\eps)^2/\eps^2)$.
    
\end{proof}

Again, we divide the samples selected in Line~\ref{en:line1-tree} in $\TreeTesting$ into two distinct groups for analysis purposes.
\begin{itemize}
\item The first group consists of a set $\bS = \{ \bi_1,\dots, \bi_{s/2} \}$ of the first $O(\log(1/\eps)/\eps)$ samples. Again, we use the set $\bS$ to define a ``skeleton partition'' (Definition~3.5 in~\cite{PR01} and Definition~\ref{def:skeleton-separator-tree} below).
\item The second group consists of the remaining $O(\log(1/\eps) / \eps)$ samples, which are divided into pairs $(\bi_{\ell}, \bi_{\ell+1})$ of indices. The violations we find are formed by a pair $(\bi_{\ell}, \bi_{\ell+1})$ among the second group which violates a constraint imposed by the skeleton partition of $\bS$ in the first group.
\end{itemize}

As before, the first group of samples $\bS$ puts structural constraints on the remaining points by partitioning them into equivalence classes. The notions of separator corruptions and easy-to-detect corruptions as well as parts follow in a very similar fashion.

\begin{Definition}[Skeleton Partition and Equivalence Classes]\label{def:skeleton-separator-tree}
    Let $M \in \calC$ and $S \subset [n]$. We say that $S$ is consistent if the $|S| \times |S|$ submatrix $M_{|S \times S}$ encodes a tree metric. For a consistent set $S$, we let:
    \begin{itemize}
        \item \emph{\textbf{Consistent Points}}. A point $j \in [n] \setminus S$ is a consistent point if $S \cup \{j\}$ is consistent. 
        \item \emph{\textbf{Skeleton Partition}}. A skeleton partition of $S$ is a partition $P_1,\dots, P_{\ell}$ of consistent points in $[n] \setminus S$ where two points $i,j\in[n]\setminus S$ are in the same part $P_{\ell}$, or equivalence class, if and only if $M(i,u)-M(i,v)=M(j,u)-M(j,v)$ for all $u,v\in S$.
        \item \emph{\textbf{Separator Set}}. If $i,j \in [n] \setminus S$ are in different equivalence classes, their separator set $\SEP(i,j) \subset S$ is given by
        \[ \SEP(i,j) = \left\{ (u,v) \in S^2 : M(i,u)-M(i,v)\neq M(j,u)-M(j,v) \right\}. \]
        \item \emph{\textbf{Separator Corruption}}. If $i,j \in[n]\setminus S$ are in different equivalence classes defined over $S$, the pair $(i,j)$ is a separator corruption if there exists $(u,v) \in \SEP(i,j)$ where 
        \begin{align*}
            M(i,j)+M(u,v)\neq \max\{M(i,u)+M(j,v),M(i,v)+M(j,u)\}. 
        \end{align*}
        We let $\SC(M, S) \subset ([n] \setminus S) \times ([n] \setminus S)$ denote the set of pairs that are separator corruptions.
    \end{itemize}
\end{Definition}
The equation in the Skeleton Partition definition conveys the facts that, out of the three matchings induced by $i,j,u,v$, the matching $M(i,u)+M(j,v)$ equals the matching $M(i,v)+M(j,u).$ In order for the four points to obey the four-point condition, we only need $M(i,j)+M(u,v)$ to be no more than this quantity. On the other hand, if two points $i,j$ fall into different classes, then $M(i,j)$ can be deduced exactly using the separator of pair $(i,j)$. The definition of Separator Corruption indicates that if $M(i,j)$ does not equal this deducted quantity for some $(u,v)$ in the separator set of $(i,j)$, the quadruple $\{i,j,u,v\}$ is a violation.

The idea for the algorithm is similar to the ultrametric testing algorithm. If the first group of samples $\bS$ is not consistent, then there is already a violation in the first group and we are done. If $\bS$ is consistent, on the other hand, we use $\bS$ to make a skeleton partition on the remaining points. We aim to sample a separator corruption $(\bi,\bj)$ which forms a violation with $\bS$. There is, as before, another type of corruption, the easy-to-detect corruption.


\begin{Definition}[Easy-to-Detect Corruption]
    For $M \in \calC$ and a consistent set $S\subset[n]$, let $(i,j) \in ([n] \setminus S) \times ([n] \setminus S)$ be in the same equivalence class. The pair $(i,j)$ form an easy-to-detect corruption if there exists $u,v \in S$ where
    \[ M(i,j)+M(u,v)>M(i,u)+M(j,v)=M(i,v)+M(j,u).\]
    Let $\EC(M,S) \subset ([n] \setminus S) \times ([n] \setminus S)$ be set of easy-to-detect corruptions in $M$ with respect to $S$.
\end{Definition}
Now, we formalize the idea that after sampling $\bS$, we aim to find many separator corruptions and easy-to-detect corruptions, which act as certificates that $M$ is not a tree metric. The below Lemma~\ref{lem:corruptions-tree} is an exact replicate of Lemma~\ref{lem:corruptions} in the ultrametric section. Thus, the proof of Tree Metric Testing Lemma~\ref{lem:tree-algo} assuming Lemma~\ref{lem:corruptions-tree} is exactly as the proof of Ultrametric Testing Lemma~\ref{lem:ultra-algo} using Lemma~\ref{lem:corruptions}.
\begin{lemma}\label{lem:corruptions-tree}
    Let $M \in \calC$ be $\eps$-far from $\calP^T$. Then, with probability at least $5/6$ over the draw of $\bS = \{ \bi_1,\dots, \bi_{s/2} \} \subset [n]$ where $\bi_1,\dots, \bi_{s/2} \sim [n]$ and $s = O(\log(1/\eps)/\eps)$, one of the followings holds:
    \begin{itemize}
        \item $\bS$ is not consistent;
        \item There are at least $\eps n/32$ inconsistent points in $[n] \setminus \bS$; 
        \item $|\SC(M, \bS)| + |\EC(M,\bS)| \geq \eps n^2/8.$
    \end{itemize}
\end{lemma}

\begin{proof}[Proof of Lemma~\ref{lem:tree-algo} assuming Lemma~\ref{lem:corruptions-tree}] 
The proof is exactly the same as that of Lemma~\ref{lem:ultra-algo}.
\end{proof}
To show Lemma~\ref{lem:corruptions-tree}, we re-use the ideas of categorizing different parts and analyze the dynamics between the parts.
\begin{Definition}[Easy, Versatile, and Active Parts]\label{def:parts-tree}
    Let $M\in\calC$ be $\eps$-far from $\calP^T$ and $S \subset [n]$ be consistent, and let $P_1,\dots, P_{\ell}$ denote the skeleton partition. 
    \begin{itemize}
        \item \textbf{\emph{Easy Part}}. A part $P_{\ell}$ is \emph{easy} if 
        \[ \Prx_{\bj, \bk \sim P_{\ell}}\left[(\bj,\bk) \in \EC(M, \bS) \right] \geq \frac{1}{2}.  \]
        \item \textbf{\emph{Versatile Part}}. Consider a part $P_{\ell}$ with $\beta n$ indices (for $\beta > 0$). We say $P_{\ell}$ is \emph{versatile} if the following hold:
        \begin{itemize}
            \item there exists a $(\beta n)\times(\beta n)$ matrix $M^\ast_{P_\ell}$ such that the metric on subset $S\cup P_\ell$ defined by \[
            D(i,j)=\left\{
            \begin{array}{cc}
                 M(i,j) &\text{if $i$ or $j$ or both in $S$}\\
                 M^{\ast}_{P_\ell}(i,j)&\text{if }i,j\in P_\ell 
            \end{array}
            \right.
            \] is a tree metric
            \item $\|M^\ast_{P_\ell}-M_{|P_{\ell} \times P_{\ell}}\|\leq \beta\cdot \eps n^2/2$.
        \end{itemize}
        That is, intuitively, no more than $\beta\cdot \eps n^2/2$ entries of the sub-block-matrix $M_{|P_{\ell} \times P_{\ell}}$ need to be modified so that $M_{|P_{\ell} \times P_{\ell}}$ is fixed into a tree metric consistent with $S$.
        \item \textbf{\emph{Active Part}}. A part $P_{\ell}$ is \emph{active} if it is neither easy nor versatile. We let \
        \begin{align*} 
        A(M, S) &= \left\{ (j,k) \in P_{\ell} \text{ for some active part $P_{\ell}$} \right\} \\
        \alpha(M, S) &= \Prx_{\bj \sim [n] }\left[ \bj \in P_{\ell} \text{ for some active part $P_{\ell}$}\right]
        \end{align*} 
    \end{itemize}
\end{Definition}
\begin{lemma}\label{lem:large-EC-ODC-tree}
    Let $M\in\calC$ be $\eps$-far from $\calP^T$ and $S \subset [n]$ be consistent. If $|A(M, S)|\leq \eps n^2/8$, then either there are at least $\eps n/32$ inconsistent points in $[n]\setminus S$, or $|\SC(M, S)| + |\EC(M,S)| \geq \eps n^2/8$.
\end{lemma}
\begin{proof}
    We prove the contra-positive. That is, given $M \in \calC$ and a consistent $S \subset [n]$ which satisfies (i) $|A(M, S)| \leq \eps n^2/8$, (ii) $|\SC(M,S)| + |\EC(M,S)| \leq \eps n^2/8$, and (iii) at most $\eps n / 32$ inconsistent points in $[n] \setminus S$, there exists an $n \times n$ matrix $\tilde{M} \in \calP^{T}$ that differs from $M$ on fewer than $\eps n^2$ entries. The matrix $\tilde{M}$ is constructed as follows:
    \begin{enumerate}
        \item For all $i, i' \in S$ we set $\tilde{M}(i,i')$ to $M(i,i')$.
        \item\label{en:second-change-tree} If $j \in [n] \setminus S$ is among the $\eps n /32$ inconsistent points, we let $\tilde{M}(j,k)$ be arbitrarily large for all $k\in[n]$, which corresponds to effectively removing $j$ while keeping an tree metric. The points $j \in [n] \setminus S$ which are consistent have $\tilde{M}(j,i)$ be set to $M(j,i)$ for $i \in S$.
        \item\label{en:third-change-tree} For $j,k \in [n] \setminus S$ which are consistent and lie in different equivalence classes:
        \begin{itemize}
            \item If $(j,k) \notin \SC(M,S)$, set $\tilde{M}(j,k)=M(j,k)$. 
            \item If $(j,k)\in \SC(M,S),$ find $(u,v)\in\SEP(j,k)$ and set $\tilde{M}(j,k)=\max\{M(i,u)+M(j,v),M(i,v)+M(j,u)\}-M(u,v)$. (As we will see, the specific choice of $(u,v)$ will not matter).
        \end{itemize}
        \item For $j,k\in[n]\setminus S$ which are consistent and in the same equivalence class $P_{\ell}$:
        \begin{itemize}
            \item If $P_{\ell}$ is an Easy Part, find a $u\in S$ and set $\tilde{M}(j,k)=M(u,j)+M(u,k)-\min_{j',k'\in P_\ell}\{M(u,j')+M(u,k')\}$.
            \item If $P_{\ell}$ is an Active Part, find a $u\in S$ and set $\tilde{M}(j,k)=M(u,j)+M(u,k)-\min_{j',k'\in P_\ell}\{M(u,j')+M(u,k')\}$.
            \item If $P_{\ell}$ is a Versatile Part, find the matrix $M^\ast_{P_{\ell}}$ as in Definition~\ref{def:parts-tree}, and set $\tilde{M}(j,k)=M^\ast_{P_{\ell}}(j,k)$.
        \end{itemize}
    \end{enumerate}

    We now show that $\|\tilde{M}-M\|_{0}<\eps n^2$. By the same argument as in Lemma~\ref{lem:large-EC-ODC} in ultrametric testing, $M$ and $\tilde{M}$ differ on at most
    \[ \frac{\eps n}{32} \times 2n + |\SC(M,S)| + \sum_{\substack{P_{\ell} \text{ easy}}} |P_{\ell}|^2 + |A(M,S)| + \sum_{P_{\ell} \text{ versatile}} |P_{\ell}| \cdot \eps n / 2\leq 15/16\cdot \eps n^2 \]

    entries. With similar ideas as in Lemma~\ref{lem:large-EC-ODC}, any quadruple $\{i,j,k,l\}$ does not form a violation in $\tilde{M}$, which makes $\tilde{M}$ a tree metric. Since $M$ is $\eps$-far from $\calP^T$, we must have either at least $\eps n/32$ inconsistent indices in $[n]\setminus S$ or $|\SC(M,S)|+|\EC(M,S)|\geq \eps n^2/8$.

\end{proof}

\begin{lemma}\label{lem:AS-large-tree}
    Suppose $M\in\calC$ is $\eps$-far from $\calP^T$. With probability at least $5/6$, over the draw of $\bS\subset[n]$ of size at least $\tilde{\Omega}(1/\eps)$, $|A(M,\bS)|\leq \eps n^2/8$.
\end{lemma}
\begin{proof}[Proof of Lemma~\ref{lem:corruptions-tree} assuming Lemma~\ref{lem:AS-large-tree}]
    The proof is exactly the same as the proof of Lemma~\ref{lem:corruptions}.
\end{proof}

\begin{lemma}\label{lem:active-entry-reduce_AS-tree}
    Suppose $M\in\calC$ is $\eps$-far from $\calP^T$. Suppose $S=\{i_1,...,i_{j-1}\}\subset[n]$. Over the randomness of the index $\bi_j$ sampled from $[n]$, $\Ex_{\bi_j}[|A(M,S\cup\{\bi_{j}\})|]\leq |A(M,S)|\cdot\exp(-\frac{\eps}{16}).$
\end{lemma}
\begin{proof}[Proof of Lemma~\ref{lem:AS-large-tree} assuming Lemma~\ref{lem:active-entry-reduce_AS-tree}]
    The proof is exactly the same as the proof of Lemma~\ref{lem:AS-large}.
\end{proof}

\begin{lemma}\label{lem:sample-active-tree}
    Suppose $M\in\calC$ is $\eps$-far from $\calP^T$. Suppose $S\subset[n]$. Over the randomness of the next sampled index $\bi_j$, if $\bi_j$ is sampled from an Active Part $P$ with size $\beta n$, 
    \[
    \Ex_{\bi_j}[|A(M,S\cup\{\bi_j\})||\bi_j\in P]\leq |A(M,S)|-\beta\cdot \eps n^2/16.
    \]
\end{lemma}
\begin{proof}[Proof of Lemma~\ref{lem:active-entry-reduce_AS-tree} assuming Lemma~\ref{lem:sample-active-tree}]
The proof is exactly the same as the proof of Lemma~\ref{lem:active-entry-reduce_AS}.
\end{proof}

The chain of implications between lemmas and the proofs of the lemmas are all the same as in the ultrametric case, with only the proof of the last Lemma~\ref{lem:sample-active-tree} being different. We now present this proof.
\begin{proof}[Proof of Lemma~\ref{lem:sample-active-tree}]

    Let $B$ be the sub-block-matrix $M_{|P\times {P}}$ representing the part $P$. Suppose the $(j,k)^{th}$ entry of block $B$ gives the distance $M(i_j,i_k)$ between indices $i_j$ and $i_k$. We fix an arbitrary sampled point $u\in S$ and define a $\beta n\times \beta n$ masking matrix $\mathrm{MASK}$ where the $(j,k)^{th}$ entry of $\mathrm{MASK}$ equals $M(u,i_j)+M(u,i_k)$. That is,
\[
\renewcommand\arraystretch{1.3}
B=\mleft[
\begin{array}{cccc}
  M(i_1,i_1) & M(i_1,i_2) &\cdots &M(i_1,i_{\beta n})\\
   M(i_2,i_1) & M(i_2,i_2) &\cdots &M(i_2,i_{\beta n})\\
   M(i_3,i_1)& M(i_3,i_2) &\cdots &M(i_3,i_{\beta n})\\
  
  \vdots & \vdots & \ddots  &\vdots\\
   M(i_{\beta n},i_1) & M(i_{\beta n},i_2)  &\cdots &M(i_{\beta n},i_{\beta n})\\
\end{array}
\mright]
,
\]
\[
\renewcommand\arraystretch{1.3}
\mathrm{MASK}=\mleft[
\begin{array}{cccc}
  M(u,i_1)+M(u,i_1) & M(u,i_2)+M(u,i_1)  &\cdots &M(u,i_{\beta n})+M(u,i_1)\\
   M(u,i_1)+M(u,i_2) & M(u,i_2)+M(u,i_2)  &\cdots &M(u,i_{\beta n})+M(u,i_2)\\
   M(u,i_1)+M(u,i_3) & M(u,i_2)+M(u,i_3)  &\cdots &M(u,i_{\beta n})+M(u,i_3)\\
  
  \vdots & \vdots & \ddots  &\vdots\\
   M(u,i_1)+M(u,i_{\beta n}) & M(u,i_2)+M(u,i_{\beta n})  &\cdots &M(u,i_{\beta n})+M(u,i_{\beta n})\\
\end{array}
\mright]
.
\]
Lemma~\ref{lem:sample-active} in the ultrametric testing section considers the values in the block matrix $B$. Here, instead, we consider the values in the square matrix $(B-\mathrm{MASK})$. The motivation is that, when we sample some index $\bi_j$ from the part $P$, other indices in $P$ are partitioned into equivalence classes according to the values in the $j^{th}$ row/column of $(B-\mathrm{MASK})$. In particular, for indices $i_a,i_b$, 
\begin{align*}
&i_a,i_b\text{ are separated into two equivalence classes according to }S\cup\{\bi_j\}\\
\Leftrightarrow \,& \exists\, w\in S, M(\bi_j,i_a)+M(w,i_b)\neq M(w,i_a)+M(\bi_j,i_b)\\
\Leftrightarrow\,&M(\bi_j,i_a)+M(w,i_b)+M(u,i_b)\neq M(w,i_a)+M(\bi_j,i_b)+M(u,i_b)=M(w,i_b)+M(\bi_j,i_b)+M(u,i_a)\\
&\text{(equality follows from the fact that }i_a,i_b\text{ are in the same equivalence class under }S)\\
\Leftrightarrow\,&M(\bi_j,i_a)+M(u,i_b)\neq M(\bi_j,i_b)+M(u,i_a)\\
\Leftrightarrow \,&M(\bi_j,i_a)-M(u,\bi_j)-M(u,i_a)\neq M(\bi_j,i_b)-M(u,\bi_j)-M(u,i_b)\\
\Leftrightarrow \,&\text{ the }(\bi_j,i_a)\text{ and }(\bi_j,i_b)\text{ entry of the matrix }B-\mathrm{MASK}\text{ are different.}
\end{align*}

Reusing the argument idea of Lemma~\ref{lem:sample-active}, among the entries in $B\setminus\EC(M,S)$, suppose the set of the values these entries take on in $(B-\mathrm{MASK})$ is $V=\{v_1,v_2,...,v_k\}$, and further suppose $v_1$ is the value that most entries take on. Suppose $V'=\{v_{k+1},...,v_{l}\}$ is the set of values the easy-to-detect corrupted entries in $B$ take on in the matrix $(B-\mathrm{MASK})$. We first show that any value in $V'$ is larger than any value in $V$, so as a result $\{v_1,...,v_k,v_{k+1},...,v_l\}$ is indeed a set of non-repetitive values.
\begin{claim}
Suppose $v'\in V'$ and $v\in V$. Then $v'>v$.
\end{claim}
\begin{proof}
    Suppose the entry $(i_a,i_b)$ in $(B-\mathrm{MASK})$ takes on the value $v'$, which implies that $M(i_a,i_b)\in \EC(M,S)$ and $v'=M(i_a,i_b)-M(u,i_a)-M(u,i_b)$. Suppose the entry $(i_c,i_d)$ in $(B-\mathrm{MASK})$ takes on the value $v$, which implies that $M(i_c,i_d)$ is not in $\EC(M,S)$. Moreover, $v=M(i_c,i_d)-M(u,i_c)-M(u,i_d)$. Since $M(i_a,i_b)$ is an easy-to-detect corruption, there exist some $x,y\in S$ such that $M(i_a,i_b)+M(x,y)>M(i_a,x)+M(i_b,y)$. On the other hand, as $M(i_c,i_d)$ is not an easy-to-detect corruption, for the pair $x,y$, $M(i_c,i_d)+M(x,y)\leq M(i_c,x)+M(i_d,y).$

    Following the above facts, we have the inequalities
    \begin{align}
        v'=&M(i_a,i_b)-M(u,i_a)-M(u,i_b)\\
        >&M(i_a,x)+M(i_b,y)-M(x,y)-M(u,i_a)-M(u,i_b)\\
        =&M(i_c,i_d)+M(i_a,x)+M(i_b,y)-(M(x,y)+M(i_c,i_d))-M(u,i_a)-M(u,i_b)\\
        \geq&M(i_c,i_d)+M(i_a,x)+M(i_b,y)-(M(i_c,x)+M(i_d,y))-M(u,i_a)-M(u,i_b)\\
        =&M(i_c,i_d)+M(i_a,x)+M(i_b,y)-(M(i_c,x)+M(i_a,u))-(M(i_d,y)+M(i_b,u))\\
        =&M(i_c,i_d)+M(i_a,x)+M(i_b,y)-(M(i_c,u)+M(i_a,x))-(M(i_d,u)+M(i_b,y))\\
        =&M(i_c,i_d)-M(i_c,u)-M(i_d,u)=v,
    \end{align}
    where (2) and (4) follow from the above facts on easy-to-detect corruptions, (6) follows from the definition of $i_c,i_d,i_a,i_b$ being in the same part, and the other equations are merely rearrangements of terms.
\end{proof}
    By exactly the same argument as in Lemma~\ref{lem:sample-active}, if we let $r_1$ to denote the number of entries taking on value $v_1$, then over the randomness of sampling an index $\bi_j$ from $P$, the expected number of pairs of indices in $P$ that are separated is at least $((\beta n)^2-r_1)/8$. Moreover, $|A(M,S)|-|A(M,S\cup\{\bi_j\})|$ is at least the number of pairs being separated. What we left in this proof is to find a bound for $r_1$.

    Recall that in Lemma~\ref{lem:sample-active}, we argued that there is a block matrix encoding ultrametric with all entries taking on value $v_1$. The same argument applies for $(B-\mathrm{MASK})$. If we set a $(\beta n)\times (\beta n)$ matrix to be an all-$v_1$ matrix, then after un-masking this matrix, we obtain a new block matrix $B'$ which is a tree metric and consistent with $S$. Notice that this matrix only differs with $B$ on $(\beta n)^2-r_1$ entries. 
    \begin{claim}
        Since $P$ is an Active Part, $\left(\sum_{\tau=2}^lr_\tau\right)=(\beta n)^2-r_1\geq \beta \cdot \eps n^2/2$.
    \end{claim}
    \begin{proof}
        We prove this by contradiction. Suppose $\left(\sum_{\tau=2}^lr_\tau\right)=(\beta n)^2-r_1< \beta \cdot \eps n^2/2$. Create a $\beta n\times\beta n$ square matrix $M^\ast_P$, which is the sum of $\mathrm{MASK}$ and an all-$v_1$ matrix of dimension $\beta n \times \beta n$. Then the metric on subset $S\cup P$ defined by 
        \[
        D(i,j)=\left\{ 
        \begin{array}{cc}
            M(i,j) & \text{if $i$ or $j$ or both in $S$} \\
             M^\ast_P(i,j)& \text{if $i,j\in P$} 
        \end{array}
        \right.
        \]
        is a tree metric. Moreover, $\|M^\ast_P-B\|\leq \sum_{\tau=2}^lr_\tau< \beta \cdot \eps n^2/2$. Therefore, $M^\ast_P$ satisfies all conditions that make $P$ a Versatile Part. However, by assumption, $P$ is Active, which leads to a contradiction.
    \end{proof}
    As a summary, we have the sequence of inequalities $$8\cdot \left(|A(M,S)|-\Ex_{\bi_j}[|A(M,S\cup\{\bi_j\})||\bi_j\in P]\right)\geq(\beta n)^2-r_1\geq  \beta\cdot\epsilon n^2/2.$$ 
\end{proof}
\newcommand{\Ds}{\mathcal{D}_s}
\newcommand{\Dq}{\mathcal{D}_q}
\newcommand{\cn}{\nu}

\section{Ultrametric and Tree Metric Testing Lower Bound---Theorem~\ref{thm:trees-lb}}\label{sec:tree-lb}

In this section, we present two distributions $\Ds$ and $\Dq$, both supported on $n\times n$ square matrices and $\eps$-far from both ultrametrics and tree metrics. Any testing algorithm that samples $o(1/\eps)$ indices will not find any violation to either ultrametric or tree metric property in $\bM\sim \Ds$ with probability at least $2/3$; any testing algorithm that makes $o(1/\eps^{4/3})$ queries will not find any violation to either ultrametric or tree metric property in a draw $\bM\sim \Dq$ with probability at least $2/3$.

\subsection{Sample Complexity Lower Bound Distribution $\Ds$}
\paragraph{Description of $\Ds$.} We describe a single $n \times n$ matrix $M$, which we will then randomize by re-shuffling indices in $n$. Divide the $n$ indices into two groups, $G=\{x_1,x_2,...,x_r\},$ $B=\{y_1,...,y_s\}$, where $|G|=r=n-\lfloor\epsilon n\rfloor$ and $|B|=s=\lfloor\epsilon n\rfloor$. Set the diagonal entries of $M$ to 0 and the remaining entries as follows, 
\[
M(x_i,x_j)=2n,M(x_i,y_j)=M(y_j,x_i)=2n+i,M(y_i,y_j)=2n,\,\,\forall\,x_i,x_j\in G,y_i,y_j\in B.
\]

\begin{claim}\label{cl:tree-vio}
    Matrix $M$ constructed as above satisfies the below properties.
    \begin{itemize}
        \item Suppose the triple $\{a,b,c\}$ is a violation of ultrametric property in $M$. Then after re-naming $a,b,$ and $c$, $a,b\in G,c\in B$.
        \item Suppose the quadruple $\{a,b,c,d\}$ is a violation of the tree property in $M$. Then after re-naming $a,b,c,$ and $d$, $a,b,c\in G,d\in B$.
    \end{itemize}
\end{claim}
\begin{proof}
    For the first item, suppose $a,b,c\in G$, then all three pairwise distances equal $2n$. If one index is in $G$ and another two are in $B$, suppose the index in $G$ is $x_i$ for some $i\in[r]$. Then the three pairwise distances equal $\{2n,2n+i,2n+i\}$. Lastly, if all three indices are in $B$, then all three pairwise distances equal $2n$. None of the above triples are a violation of the ultrametric property. 

    Similarly, for the second item, if all four indices are in $G$, then all three induced matchings equal $2n+2n$. If two indices are in $G$, which we denote as $x_i,x_j$ for some $i,j\in[r]$, and two other indices in $B$, then the three induced matchings equal $2n+2n,(2n+i)+(2n+j),(2n+j)+(2n+i)$. If only one index is in $G$, which we denote as $x_i$ for some $i\in[r]$, and three other indices are in $B$, all three induced matchings equal $2n+(2n+i)$. Lastly, if all four indices are in $B$, all three induced matchings equal $2n+2n$. None of the above quadruples are violations of the tree metric four-point-condition.
\end{proof}

\begin{claim}
    The matrix $M$ is $\Omega(\epsilon)$-far from an ultrametric and $\Omega(\epsilon)$-far from a tree metric.
\end{claim}
\begin{proof}
First, the $n\times n$ matrix $\tM$ with every entry equal to $2n$ encodes an ultrametric and a tree metric. Moreover, $\|M-\tM\|_0=2\lfloor\epsilon n\rfloor(n-\lfloor\epsilon n\rfloor)=\Omega(\eps n^2)$.

On the other hand, for $i<j\in[r],l\in [s]$, all triples of the form $\{x_i,x_j,y_l\}$ violate the ultrametric property, since $M(x_i,x_j)=2n<M(x_i,y_l)=2n+i<M(x_j,y_l)=2n+j$. There are ${n-\lfloor\epsilon n\rfloor\choose 2}\lfloor\epsilon n\rfloor=\Omega(\epsilon n^3)$ such violating triples. Each pair $(x_i,y_l)$ or $(y_l,x_i)$ participates in $n-\lfloor\epsilon n\rfloor$ many triples; each pair $(x_i,x_j)$ participates in $\lfloor\epsilon n\rfloor$ such triples. Suppose matrix $M'$ encodes an ultrametric and $S'$ is the set of entries $M$ and $M'$ differ upon. Then, all $\Omega(\eps n^3)$ violating triples need to be covered by some pairs of indices in $S'$, which indicates $\|M'-M\|_0=|S'|= \Omega(\epsilon n^2).$ 

Similarly, for $i,j,k\in[r],l\in[s]$, any quadruple of the form $(x_i,x_j,x_k,y_l)$ violates the four-point condition for tree metric property. There are $\Omega(\epsilon n^4)$ such quadruples. Each pair $(x_i,y_l)$ participates in at most $\Omega(n^2)$ such quadruples, and each pair $(x_i,x_j)$ participates in at most $\Omega(\epsilon n^2)$ such quadruples. Suppose matrix $M''$ encodes a tree metric and $S''$ is the set of entries $M$ and $M''$ differ upon. Then, all $\Omega(\eps n^4)$ violating quadruples need to be covered by some pairs of indices in $S''$, which indicates $\|M''-M\|_0=|S''|=\Omega(\epsilon n^2).$ 
\end{proof}

We let $\bM \sim \Ds$ be obtained from $M$ by re-ordering rows and columns according to a uniformly random permutation $\bpi$. By Yao's minimax principle, it suffices to rule out any deterministic ultrametric testing algorithm for finding violating triples and any deterministic tree metric testing algorithm for finding violating quadruples in a draw $\bM \sim \Ds$.

\begin{lemma}
    Consider any deterministic non-adaptive testing algorithm which samples $o(1/\eps)$ indices. With probability at least $2/3$ over the draw of $\bM \sim \Ds$, there are no indices $i,j,k$ with 
    \[ \bM(i,j) > \max\{\bM(i,k), \bM(k,j)\};\]
    Moreover, with probability at least $2/3$ over the draw of $\bM \sim \Ds$, there are no indices $i,j,k,l$ with
    \[
    \bM(i,j)+\bM(k,l)>\max\{\bM(i,k)+\bM(j,l),\bM(i,l)+\bM(j,k)\}.
    \]
    For partial matrix $M'$ that contains the pairwise distance between each pair of indices in a subset $S\subset[n]$ and does not violate the ultrametric three-point-condition (or tree metric four-point-condition), $M'$ may be completed to one which is an ultrametric (or tree metric). Thus, any non-adaptive testing algorithm must sample $\Omega(1/\eps)$ indices.
\end{lemma}
\begin{proof}
    For the first claim, it suffices to show that for any testing algorithm which samples $o(1/\eps)$ indices $S$, with probability at least 2/3 over the draw of $\bM\sim\Ds$, $S\cap B=\emptyset$. Then, by Claim~\ref{cl:tree-vio}, all triples in $S$ are not violations of the ultrametric property. Thus, the probability that a violation of ultrametric property is found among the indices $S$, by a union bound, is at most
    \[
    \sum_{b\in B}\Pr_{\bpi}\left(\bpi(b)\in S)\right)=|B|\cdot \frac{|S|}{n}=o(1)
    \] when $|S|=o(1/\eps)$.

    For the second claim, the same argument as above holds. That is, by Claim~\ref{cl:tree-vio}, the probability that a violation of tree metric property is found among the indices $S$ is at most the probability that some index in $B$ fall into $S$. By the same union bound, this probability is at most $o(1)$ when $|S|=o(1/\eps)$.
\end{proof}

\subsection{Query Complexity Lower Bound Distribution $\Dq$}

Notice that any ultrametric is also a tree metric. We show below a distribution $\Dq$ such that any testing algorithm that makes $o(1/\eps^{4/3})$ queries will not find any violation of the ultrametric property in a draw $\bM\sim \Dq$ with probability at least $2/3$. Thus, any testing algorithm that makes $o(1/\eps^{4/3})$ queries will not find any violation of the tree metric property in a draw $\bM\sim \Dq$ with probability at least $2/3$.

\paragraph{Description of $\Dq$.} As before, we describe a single $n \times n$ matrix $M$, which we will then randomize by re-shuffling indices in $n$. First, let $r=\lfloor\eps n\rfloor$. We define a $r\times r$ block matrix $D$ and then use multiple block diagonal matrices $D$ to construct $M$. First, $D$ is a randomized $r\times r$ matrix with 0 on the diagonal; for $i<j\in[r],$ set entry $D(i,j)=D(j,i)$ to be 1 with probability 1/2 and 2 with probability 1/2. Partition $[n]$ into $l$ groups of indices $[n]=(\cup_{j=1}^lI_j)\cup R$ for $l=\lfloor 1/\eps\rfloor$ such that each group $I_j$ has $r$ indices, and possibly a ramainder group $R$ with less than $r$ indices. In particular, for $j\in[l]$, $I_j$ contain all integers in $((j-1)r,jr]$. For each $j\in[l]$, sample a random $r\times r$ matrix $D_j$ and set $M_{|I_j\times I_j}=D_j$. This is equivalent to letting $l$ random block matrices $D_j$ to be the block diagonal matrices of $M$. Set all other entries, i.e. the entries that are not in the block diagonal matrices $D_j$ and the entries in the remainder group $M_{|R\times R}$, to be 10.

Below is a demonstration of the matrix $M$, which has $l$ random block diagonal matrices of dimension $r\times r$; the entries that are not in the block diagonal matrices are 10.
\[
\renewcommand\arraystretch{1.3}
M=\mleft[
\begin{array}{c|c|c|c|c|c}
  D_1 & \ast &\ast &\cdots &\ast&\ast\\
  \hline
  \ast & D_2 & \ast &\cdots &\ast&\ast\\
  \hline
  \ast & \ast & D_3 &\cdots &\ast&\ast\\
  \hline
  \vdots & \vdots & \vdots &\ddots &\vdots&\vdots\\
  \hline
  \ast & \ast & \ast &\cdots &D_l&\ast\\
  \hline
  \ast & \ast & \ast &\cdots &\ast&R\\
\end{array}
\mright].
\]
\begin{claim}
    The matrix $D$ is $\Omega(1)$-far from an ultrametric. In consequence, the matrix $M$ is $\Omega(\eps)$-far from an ultrametric.
\end{claim}
\begin{proof}
    For indices $i<j<k\in[r]$, each entry $D(i,j),D(i,k),D(j,k)$ takes on value $1$ with probability $1/2$ and $2$ with probability $1/2$. Out of 8 arrangements of values that happen with equal probability, 3 of them incur a violation of the ultrametric property. Thus, with probability $3/8$ the triple violates the ultrametric property. In the matrix $D,$ there are $\Omega(r^3)$ such violating triples. Each pair of indices participates in $\Omega(r)$ such violating triples. Suppose the $r\times r$ matrix $\tilde{D}$ encodes an ultrametric; further suppose $E$ is the set of entries $\tilde{D}$ and $D$ differ on. Then, each violating triple in $D$ needs to be covered by some entry in $E$. This suggests $|E|=\Omega(r^2)$, so $D$ is $\Omega(1)$-far from ultrametric. 

    Suppose $n\times n$ matrix $\tilde{M}$ encodes an ultrametric, then $\tM$ and $M$ differ on $\Omega(r^2)$ entries on each block diagonal matrix, and there are $\Omega(1/\eps)$ block diagonal matrices. This shows that $\|M-\tM\|_0\geq \Omega(\eps n^2)$ for every $\tM\in\calP^U$. In fact, no more than $\eps n^2$ entries need to be modified for $M$ to be an ultrametric, as $\|M-M'\|_0\leq \eps n^2$ for $M'$ being a $n\times n$ matrix with each entry set to 10.
\end{proof}

\begin{claim}\label{cl:tree-vio-q}
    Suppose the triple $\{a,b,c\}$ is a violation of the ultrametric property in $M$. Then three indices are in the same index group $I_j$ for some $j\in[l]$.
\end{claim}
\begin{proof}
    We prove this by showing the contra-positive is true. Suppose less than 3 indices are in the same index group $I_j$ for any $j\in[l]$. Then out of the three distances $M(a,b),M(a,c),M(b,c)$, at least two of them equal 10. The remaining distance is no larger than 10. Thus, the ultrametric is not violated for $\{a,b,c\}$.
\end{proof}

We let $\bM \sim \Dq$ be obtained from $M$ by re-ordering rows and columns according to a uniformly random permutation $\bpi$. Again, it suffices to rule out any deterministic ultrametric testing algorithm for finding violating triples in a draw $\bM \sim \Ds$.

\begin{lemma}
    Consider any deterministic non-adaptive algorithm that makes $o(1/\eps^{4/3})$ queries. With probability at least $2/3$ over the draw of $\bM \sim \Dq$, no violation to the ultrametric can be found.
    
    Since any partial matrix that does not violate the ultrametric property may be completed to one that is ultrametric, any non-adaptive testing algorithm must make $\Omega(1/\eps^{4/3})$ queries.
\end{lemma}
\begin{proof}
    Suppose a testing algorithm samples a set of indices $S$ of size $s$. The algorithm only recognizes a violation if at least three indices in $S$ fall into the same index group $I_j$ for some $j\in[l]$. Moreover, if a subset $U\subset S$ with at least three indices fall into the same index group $I_j$, the algorithm needs to make queries that form a cycle in $U$ to recognize a violation. Formally we have the below claim.

    \begin{claim}
    Suppose for a subset of indices $U\subset[n],$ $E'$ is the set of pairs whose distances are known and $E'$ is cycle-free. That is, suppose a graph $G'$ is on vertices $U$ and $(a,b)\in U^2$ is an edge if and only if $(a,b)\in E'$, and suppose $G'$ is cycle-free. Then the known distances have no violation to ultrametric.
\end{claim}
\begin{proof}
    Define $G'$ as proposed in the claim and for each edge $(a,b)$ in $G'$, set its weight to be the known distance between $a,b$. Consider the partial matrix $M'_{|U\times U}$ whose entries are the known distances in $E'$. There exists a way to complete $M'_{|U\times U}$ to another matrix $M_{|U\times U}$ which encodes an ultrametric. Namely, first, we identify all disconnected components in $G'$, and between each pair of components, find a vertex in each component and set their edge weight to be an arbitrary positive value. Now $G'$ is a tree. Then, for every non-edge $(a,b)\notin E'$, find the unique simple path from $a$ to $b$ in the graph $G'$ and set $M_{|U\times U}(a,b)$ to be the weight of the heaviest edge on the path.
\end{proof}
Upon sampling the indices $S$, the testing algorithm makes queries among a subset of pairs $E\subset S^2$. Create a graph $G$ whose vertex set is $S$ and edge set $E$. Let $\calC(G)$ be the set of cycles in $G$. For a cycle $C\subset[n]$, let $\mathbb{I}(C)$ denote the indicator variable in $\{0,1\}$ with
\[
\mathbb{I}(C)\qquad\iff\qquad \Bigl\{
\begin{array}{c}
     C\in\calC(G),  \\
     \exists\, j\in[l],\,\forall\, v\in C,\,v\in \bpi(I_j)
\end{array}
\Bigr\}
\]
Then, the probability that a violation of ultrametric is detect by the algorithm, by a union bound, is at most 
\begin{equation}\label{eq:lb-prob}
\sum_{C\in\calC(G)}\mathbb{I}(C)\leq\sum_{k=3}^\infty\sum_{\substack{C\in\calC(G),\\|C|=k}}2\eps^{k-1}
\end{equation}
where the inequality follows from when cycle $C$ has $k$ indices, all $k$ indices fall into the same index group with probability $\lfloor1/\eps\rfloor\cdot{\lfloor\eps n\rfloor\choose k}/{n\choose k}\leq2\eps^{k-1}$ as there are $\Omega(1/\eps)$ index groups of $\lfloor\eps n\rfloor$ indices. An algorithm aims to pick a query graph structure to maximize this quantity. The following claim shows that among the sampled indices $S$, making a ``clique-like" query graph structure is optimal for a testing algorithm.

\begin{lemma}\label{lem:clique-optimal}
Suppose a testing algorithm will make $m$ queries. To maximize the quantity~\ref{eq:lb-prob}, the algorithm should select $s$ indices such that ${s\choose2}\geq m>{s-1\choose 2}$ and make $m$ queries on the graph with these $s$ indices. That is, a graph with $m$ edges that is closest to a clique maximizes the quantity.
\end{lemma}

Assuming Lemma~\ref{lem:clique-optimal}, with $m$ queries, the probability that a violation to ultrametric is detected is maximized when the algorithm samples $s$ indices such that ${s\choose2}=m$ and makes $m$ queries on the pairwise distances. If an algorithm samples a set $S$ of $o(1/\eps^{2/3})$ indices and query pairwise distances, then a violation is detected when three indices in $S$ fall into the same index group. For each triple, all three indices fall into the same group with probability at most $2\eps^2$. Thus, by a union bound, the probability that the algorithm detects a violation is at most
\[
\sum_{i,j,k\in S}\Pr\left(\exists\, j\in[l],\,i,j,k\in\bpi(I_j)\right)\leq o(1/\eps^2)\cdot 2\eps^2=o(1)
\]
when $|S|=o(1/\eps^{2/3})$. Such clique-like query structure makes $o(1/\eps^{4/3})$ queries, and for any other query structure, by Lemma~\ref{lem:clique-optimal}, the probability of detecting a violation is no larger than $o(1)$. This shows that any non-adaptive testing algorithm must make $\Omega(1/\eps^{4/3})$ queries to detect a violation of ultrametric property with high probability.

\begin{proof}[Proof of Lemma~\ref{lem:clique-optimal}]
Given the choice to make $m$ queries, an algorithm wants to maximize the quantity $\sum_{k=3}^\infty\epsilon^{k-1}\cdot|C_k(G)|$ where $C_k(G)$ denotes the set of cycles of length $k$ in the query graph $G$. Let $s$ be an integer defined as in the statement. We start with an arbitrary graph $G$ with $s'>s$ vertices and $m$ edges, and show that $G$ can be modified to $G'$ with strictly less vertices such that $\sum_{k=3}^\infty\epsilon^{k-1}\cdot|C_k(G')|>\sum_{k=3}^\infty\epsilon^{k-1}\cdot|C_k(G)|$. This graph modification is done by merging two vertices and adding edges between non-adjacent vertices.

First, we modify $G=(V,E)$ such that its diameter is at most 2. Suppose initially the diameter of $G$ is larger than 2. Then, there exists two non-adjacent vertices $x,y$ such that the neighbor of $x$, denoted by $N_x(G)$, is disjoint from that of $y$, $N_y(G)$. Create a new graph $G'$ on the vertex set $(V\cup\{z\})\setminus\{x,y\}$ and an edge set $E':$
\[
E'=\{(u,v):(u,v)\in E,u,v\notin\{x,y\}\}\cup\{(v,z):(v,x)\in E\}\cup\{(v,z):(v,y)\in E\}.
\]
The following statements hold: 
\begin{itemize}
    \item For $C\in\calC(G)$, if $x,y\notin C$, $C\in\calC(G').$
    \item For $C\in\calC(G)$, if $|C\cap\{x,y\}|=1$, $C\in\calC(G').$ That is, any cycle in $G$ that only contains one of $x,y$ still remains in $G'$.
    \item For $C\in \calC(G)$,if $x,y\in C$, then $C$ must be of the form $\{x\rightarrow v_1\rightarrow\cdots\rightarrow v_i\rightarrow y\rightarrow u_1\rightarrow \cdots\rightarrow u_j\rightarrow x\}$ where $v_1\neq v_i$, $u_1\neq u_j$. Thus, the cycles $C_1=\{z\ra v_1\ra\cdots\ra v_i\ra z\}$ and $C_2=\{z\ra u_1\ra\cdots\ra u_i\ra z\}$ are in $\calC(G')$ and $|C_1|<|C|$ and $|C_2|<|C|$.
\end{itemize}
Therefore, $\sum_{k=3}^\infty\epsilon^{k-1}\cdot|C_k(G')|>\sum_{k=3}^\infty\epsilon^{k-1}\cdot|C_k(G)|$.

Hence, suppose that $G=(V,E)$ is a graph with $s'$ vertices with diameter at most 2. Since $s'>s$ where integer $s$ satisfies ${s\choose 2}\geq m$, we must have that at least ${s'\choose2}-m\geq{s'\choose2}-{s'-1\choose2}=s'-1$ pairs of vertices are non-adjacent in $G$. That is, $G$ has at least $s'-1$ non-edges. Suppose $(x,y)\notin E$, and suppose $|N_x(G)\cap N_y(G)|=\cn$. Notice that $\cn>0$ as $diam(G)\leq2$. Consider the effect of adding an edge $(x,y)$ into the graph as well as the effect of merging the pair of vertices $(x,y)$ into a single vertex $z$.
\begin{description}
\item[Create edge $(x,y)$: ] Define $G''$ to be a graph on vertex set $V$ and edge set $E\cup\{(x,y)\}$. Then $|\calC(G'')|>|\calC(G)|$. In particular, we gain $\cn$ new triangles from the $\cn$ common neighbors of $x,y\in G$. That is, $|C_3(G'')|=|C_3(G)|+\cn$ so $|\calC(G'')|\geq|\calC(G)|+\cn$.
\item[Merge vertices $(x,y)$: ] Define $G''$ to be a graph on vertex set $(V\cup\{z\})\setminus\{x,y\}$ and edge set $E''$, where
\[
E''=\{(u,v):(u,v)\in E,u,v\notin\{x,y\}\}\cup\{(v,z):(v,x)\in E\}\cup\{(v,z):(v,y)\in E\}.
\] For $C\in\calC(G)$ and $|C\cap\{x,y\}|\leq1,$ $C\in\calC(G'')$; for $C\in\calC(G)$, $x,y\in C$, and $|C|\geq5$, $C$ corresponds to at least one cycle $\tilde{C}\in\calC(G'')$ where $|\tilde{C}|<5$. For $C\in\calC(G)$, $x,y\in C$, and $|C|=4$, $C$ is of the form $\{x\ra v\ra y\ra u\ra x\}$. Such cycles are vanishing due to the merge of $(x,y).$ Moreover, for triangles of the form $\{x,u,v\},\{y,u,v\}\in C_3(G)$ for some $u,v\in V$, $\{z,u,v\}\in C_3(G'')$: after the merge, two triangles become one. These types of triangles are the only affected ones. Moreover, since $|N_x(G)\cap N_y(G)|=\cn$, $|E''|=|E|-\cn$. In summary, among the $\cn$ common neighbors in $N_x(G)\cap N_y(G)$, if $p$ pairs of them are adjacent, then merging $(x,y)$ causes $p$ triangles vanishing, ${\cn\choose2}$ cycles of length 4 vanishing, but the number of edges decrease by $\cn$.
\end{description}

Now we will show how to create $G'$ base on $G$ to increase the objective function. Suppose the set of non-edges in $G$ is 
\[
(x_1,y_1),\ldots,(x_\alpha,y_\alpha),\,\,\forall \,i\in[\alpha],\,x_i,y_i\in V,\,(x_i,y_i)\notin E.
\]
Recall that $\alpha\geq s'-1$. Each pair $(x_i,y_i)\notin E$ is associated with two quantities
\begin{align*}
\cn_i&:=|N_{x_i}(G)\cap N_{y_i}(G)|,\\
p_i&:=|\{(u,v)\in E:u,v\in (N_{x_i}(G)\cap N_{y_i}(G))\}|.
\end{align*}
That is, $\cn_i$ is the number of common neighbors of $x_i,y_i$, and $p_i$ is the number of pairs of common neighbors that are adjacent. Notice that $p_\alpha\leq {\cn_\alpha\choose 2}.$ We assume the pairs of non-edges are sorted in the way that $\cn_1\geq \cn_2\geq \cdots\geq \cn_\alpha$. The construction of $G'$ is done by merging a single pair $(x_\alpha,y_\alpha)$ and adding edges to some of the previous non-adjacent pairs.

By the previous case analysis, merging $(x_\alpha,y_\alpha)$ into one single new vertex $z$ causes $p_\alpha$ triangles and ${\cn_\alpha\choose2}$ cycles of length 4 vanishing; but we also spares $\cn_\alpha$ edges that can be added to elsewhere in the graph. Let $G''=(V'',E'')$ denote the new graph with $x_\alpha,y_\alpha$ merged into a new vertex $x$. ($G''$ is defined formally in Claim~\ref{cl:merge_v} below.) Then $|C_3(G'')|\geq|C_3(G)|-p_\alpha\geq |C_3(G)|-{\cn_\alpha\choose 2}$ and $|C_4(G'')|\geq|C_4(G)|-{\cn_\alpha\choose 2}$. Moreover, for $j>4$, $|C_j(G'')|\geq|C_j(G)|$. In terms of the number of edges, $|E''|=|E|-\cn_\alpha$.

A natural next step is to create $\cn_\alpha$ edges between some of the non-edges $(x_1,y_1),\ldots,(x_{\alpha-1},y_{\alpha-1})$ and make some of them adjacent. However, note that a by-product of merging $(x_\alpha,y_\alpha)$ into a new single vertex $z$ in $G''$ is that, some of the non-edges $(x_1,y_1),\ldots,(x_{\alpha-1},y_{\alpha-1})\in G$ become adjacent in $G''$. Consider the possible scenario where two non-edges $(x_i,y_i)$ and $(x_\alpha,y_\alpha)$ in $G$ share one same endpoint $x_i=x_\alpha$ and the pair $(y_\alpha,y_i)\in E$ is adjacent in $G$. Thus, the non-edge $(x_i,y_i)=(x_\alpha,y_i)\in G$ corresponds to $(z,y_i)\in G''$ when merging $x_\alpha$ and $y_\alpha$ as a new vertex $z\in G''$. In this case, it makes no sense to use the $\cn_\alpha$-edge-budget to create an edge between $(x_i,y_i)$ in $G''$, as the vertex $x_i=x_\alpha$ corresponds to $z$ in $G''$ which is already adjacent to $y_i$. Below, we give an upper bound on the decrease in the non-edges after the merge.

\begin{claim}\label{cl:merge_v}
    Define $G''$ to be a graph on vertex set $V''=(V\cup \{z\})\setminus\{x_\alpha,y_\alpha\}$ and edge set $E''$ where
    \[
    E''=\{(u,v):(u,v)\in E,u,v\notin\{x_\alpha,y_\alpha\}\}\cup\{(v,z):(v,x_\alpha)\in E\}\cup\{(v,z):(v,y_\alpha)\in E\}.
    \]
    Then there are at least $\cn_\alpha$ non-edges in $G''$.
\end{claim}
\begin{proof}
Suppose non-edge $(x_i,y_i)\notin E$ in $G$ becomes adjacent in $E''$. Then we must be in one of the below two cases 
\[
\Bigl\{x_i=x_\alpha\text{ and }(y_\alpha,y_i)\in E\Bigr\} \qquad \text{or}\qquad \Bigl\{y_i=y_\alpha \text{ and }(x_\alpha,x_i)\in E\Bigr\}.
\]

Consider the former case. Since $(y_\alpha,y_i)\in E$, $y_i$ was not adjacent to $x_\alpha=x_i$ as $(x_i,y_i)\notin E$. That is, $y_i\in N_{y_\alpha}(G)\setminus N_{x_\alpha}(G)$. Thus, $|\{(y_\alpha,y_i)\in E:y_i\in V\}|\leq|N_{y_\alpha}(G)\setminus N_{x_\alpha}(G)|$. Similarly, $|\{(x_\alpha,x_i)\in E:x_i\in V\}|\leq|N_{x_\alpha}(G)\setminus N_{y_\alpha}(G)|$. In summary, $|\{(x_i,y_i)\notin E,(x_i,y_i)\in E''\}|\leq |N_{y_\alpha}(G)\setminus N_{x_\alpha}(G)|+|N_{x_\alpha}(G)\setminus N_{y_\alpha}(G)|\leq s'-2-\cn_\alpha$. There are $\alpha-1$ non-edges $(i_1,j_1),\ldots,(i_{\alpha-1},j_{\alpha-1})$ in $G$. Thus, in $G''$, there are at least $\alpha-1-(s'-2-\cn_\alpha)=\alpha-s'+1+\cn_\alpha\geq \cn_\alpha$ non-adjacent pairs. 
\end{proof}

The above argument shows that there are still at least $\cn_\alpha$ pairs of vertices in $G'$ that remain to be non-adjacent, and the $\cn_\alpha$ spare edges can be added to $\cn_\alpha$ of the non-adjacent pairs. After renaming, let these $\cn_\alpha$ non-adjacent pairs be 
\[
E^{(NA)}=\{(x_i,y_i)\notin E'':i\in[\cn_\alpha],x_i,y_i\in V''
\}.
\]
Define graph $G'$ on the vertex set $V'=V''=(V\cup \{z\})\setminus\{x_\alpha,y_\alpha\}$ and edge set $E':$ 
    \begin{align*}
    E'=&\{(u,v):(u,v)\in E,u,v\notin\{x_\alpha,y_\alpha\}\}\cup\{(v,z):(v,x_\alpha)\in E\}
    \\&\cup\{(v,z):(v,y_\alpha)\in E\}\cup\{(x,y):(x,y)\in E^{(NA)}\}.
    \end{align*}

    Notice that, for each new edge $(x_i,y_i)\in E^{(NA)}$, $N_{x_i}(G')\cap N_{y_i}(G')=(((N_{x_i}(G)\cap N_{y_i}(G)))\setminus\{x,y\})\cup\{z\}$ if either $x$ or $y\in (N_{x_i}(G)\cap N_{y_i}(G))$; if $x,y\notin (N_{x_i}(G)\cap N_{y_i}(G))$, then $N_{x_i}(G')\cap N_{y_i}(G')=N_{x_i}(G)\cap N_{y_i}(G)$. In both cases, $|N_{x_i}(G')\cap N_{y_i}(G')|\geq \cn_\alpha-1$. As a consequence, for each such pair $(x_i,y_i)\in E^{(NA)},$ adding this edge in $G'$ creates at least $\cn_\alpha-1$ new triangles in $G'$. In total, adding $\cn_\alpha$ spare edges creates at least $\cn_\alpha\cdot (\cn_\alpha-1)$ new triangles. Thus, $|C_3(G')|\geq |C_3(G'')|+\cn_\alpha\cdot (\cn_\alpha-1)$ and the number of cycles with larger length can only increase.

    In summary, from graph $G$ to $G''$, we have 
    \begin{align*}
    |C_3(G'')|&\geq |C_3(G)|-\cn_\alpha(\cn_\alpha-1) 2,
    \\|C_4(G'')|&\geq|C_4(G)|-\cn_\alpha(\cn_\alpha-1) 2,
    \\|C_j(G'')|&\geq|C_j(G)|,\,\forall j>4,
    \\|E''|&=|E|-\cn_\alpha.   
    \end{align*}
    From graph $G''$ to $G'$, we have 
    \begin{align*}
    |C_3(G')|&\geq |C_3(G'')|+\cn_\alpha\cdot (\cn_\alpha-1),
    \\|C_j(G')|&\geq|C_j(G'')|,\,\forall j\geq 4,
    \\|E'|&=|E''|+\nu_\alpha.
    \end{align*}
    Thus, $|C_3(G')|\geq |C_3(G)|+{\nu_\alpha\choose2}$, $|C_4(G')|\geq |C_4(G)|-{\nu_\alpha\choose2}$, $|C_j(G')|\geq |C_j(G)|$ for all $j>4$ and the number of larger cycles in $G'$ is at least that of $G$. Recall that $\cn_\alpha>0$. This shows that, with the same number of edges $m$, $\sum_{k=3}^\infty\epsilon^{k-1}\cdot|C_k(G')|>\sum_{k=3}^\infty\epsilon^{k-1}\cdot|C_k(G)|$.
\end{proof}

\end{proof}

\section*{Acknowledgement}
Erik Waingarten would like to thank the support from the National Science Foundation (NSF) under Grant No. CCF-2337993.

\appendix

\section{Definitions of Metric Spaces}

\begin{Definition}[Metric Space] 
A metric space defined over $[n]$ is specified by a distance function $d \colon [n]\times [n] \to \R_{\geq 0}$ which satisfies:
\begin{itemize}
\item $d(i,j)=0$ if and only if $i = j$, and $d(i,j) = d(j,i)$ for all $i, j \in [n]$.
\item \emph{\textbf{Triangle Inequality}}: for all $i, j, k \in [n]$, $d(i,j)\leq d(i,k) + d(j,k)$.
\end{itemize}
Furthermore, if the only condition unsatisfied above is that $i \neq j$ satisfies $d(i,j) = 0$, then $d$ defines a \emph{pseudometric} over $[n]$. 
\end{Definition}

\begin{Definition}[Tree Metric]\label{def:tree}
A metric space $d \colon [n] \times [n] \to \R_{\geq 0}$ defines a tree metric over $[n]$ if the function $d$ may be realized in the following way:
\begin{itemize}
\item There is a weighted tree $T = (V, E)$ with weights $w \colon E \to (0, \infty)$ and an injection $\phi \colon [n] \to V$. 
\item The distance $d(i,j)$ is given by the length of the path (i.e., sum of edge weights) between $\phi(i)$ and $\phi(j)$ in $T$. 
\end{itemize}
An alternative (and equivalent) definition, due to~\cite{B75}, is that $d$ is a tree metric, in addition to being a metric, satisfies:
\begin{align*}
d(i,j) + d(k,\ell) \leq \max\left\{ d(i,k) + d(j, \ell), d(i, \ell) + d(j, k)\right\} \qquad \text{for all $i,j,k, \ell \in [n]$.}
\end{align*}
\end{Definition}

\begin{Definition}[Ultrametric]\label{def:ultra}
A metric space $d \colon [n] \times [n] \to \R_{\geq 0}$ defines an ultrametric over $[n]$ if the function $d$ may be realized in the following way:
\begin{itemize}
\item There is a rooted tree $T = (V, E)$ with a root $r \in V$ and weights $w \colon E \to (0, \infty)$ such the sum of weights from $r$ to each leaf is the same. Furthermore, there is an injection $\phi \colon [n] \to V$ which maps to the leaves of the tree. 
\item The distance $d(i,j)$ is given by the length of the path (i.e., sum of edge weights) between $\phi(i)$ and $\phi(j)$ in $T$.
\end{itemize}
An alternative (and equivalent) definition, is that $d$ defines an ultrametric if, in addition to being a metric, satisfies:
\[ d(i,j) \leq \max\{ d(i,k), d(j, k) \} \qquad \text{for all $i,j,k \in [n]$}, \]
\end{Definition}

\section{Proof of Theorem~\ref{thm:testing-metric}}

\begin{customthm}{\ref{thm:testing-metric}}[Testing Metrics---Upper Bound]
For any large enough $n \in \N$ and any $\eps \in (0,1)$, there exists a randomized algorithm which receives query access to an unknown matrix $M \in \R^{n\times n}$ and makes $O(n^{2/3} / \eps^{4/3})$ queries with the following guarantee:
\begin{itemize}
\item If $M$ defines a metric space on $[n]$, the algorithm outputs ``accept'' with probability $1$.
\item If $M$ is $\eps$-far from being being a metric, then the algorithm output ``reject'' with probability at least $2/3$.
\end{itemize}
Furthermore, the algorithm is non-adaptive (i.e., queries made do not depend on answers to prior queries).
\end{customthm}
\begin{proof}
By Lemma~\ref{lem:clean}, it suffices to show the theorem holds for $M\in \calC$. 
Note that, if $\eps < 1/n$, the claimed complexity $O(n^{2/3}/\eps^{4/3})$ is $O(n^2)$, so that we read the entire matrix. If $1/\eps>n$, run the Metric Testing Algorithm. For $M\in\calP$, both sub-routines output ``accept" with probability 1, so the algorithm outputs ``accept" if $M$ encodes a metric space. If $M$ is $\eps$-far from $\calP$, let $T$ be the set of all violating triangles in $M$ and let $I=\{i\in[n]:d_T(i)\geq \eps^{1/3}n^{4/3}/16 \}$. If $|I|\geq\eps n/4$, then sub-routine $\CheckHiDegree(M,\eps)$ outputs ``reject" with probability at least 5/6. On the other hand, if $|I|<\eps n/4$, sub-routine $\CheckViolation(M,\eps)$ outputs ``reject" with probability at least 5/6. By union bound, the algorithm outputs ``reject" with probability at least 2/3 when $M\in\calC$ is $\eps$-far from $\calP$.
\end{proof}

\section{Proofs from Section~\ref{sec:general} (Metric Testing Upper Bound)}
\begin{customlemma}{\ref{lem:exist}}
For any $\eps \in (0, 1)$, and any $M\in \calC$ which is $\eps$-far from $\calP$, there are at least $\eps n^2/6$ distinct violating triangles $\{i,j,k\}$ of $M$.
\end{customlemma}

\begin{proof}
Suppose $M$ contains less than $\eps n^2/6$ triangles which are violating for $M$. Then, we show how to modify less than $\epsilon n^2$ entries on matrix $M$ to get a matrix $M'$ which encodes a metric space over $[n]$. Let $S$ be the subset of $[n]\times[n]$ such that for all $(i,j)\in S$, $(i,j)$ does not participate in any violating triangle. Then, $|S|> n^2-\epsilon n^2$ since a triangle $\{i,j,k\}$ has $3$ (unordered) pairs of indices and hence, 6 entries in $M$. Consider the weighted undirected graph $G$ on $[n]$ which adds edges $(i,j)\in S$ with the weight $M(i,j)$. If $G$ is disconnected, add edges of the maximum weight to connect it. We then consider the matrix $M'$, where entry $M'(i,j)$ denotes the length of the shortest path between $i$ and $j$ along edges in $G$. The matrix $M'$ encodes a metric and differs with $M$ on less than $\epsilon n^2$ entries.
\end{proof}

\begin{customlemma}{\ref{lem:check-hi-degree}}[$\CheckHiDegree$ Lemma]
For $n \in \N$ and $\eps \in (1/n,1)$, there exists a randomized algorithm, $\CheckHiDegree$, which receives as input an $n\times n$ matrix $M \in \calC$ and a parameter $\eps$ and has the following guarantees:
\begin{itemize}
\item If $M \in \calP$, $\CheckHiDegree(M,\eps)$ always outputs ``accept.'' 
\item Letting $T$ be the set of violating triangles of $M$, if there are at least $\eps n/\sco$ indices $i \in [n]$ such that $d_{T}(i) \geq \ea n^{4/3}/16$, $\CheckHiDegree(M, \eps)$ outputs ``reject'' with probability at least $5/6$. 
\end{itemize} 
The algorithm is non-adaptive, taking $O(1/\eps + n^{2/3}/\ea)$ samples and using $O(n^{2/3} / \eps^{4/3})$ queries.
\end{customlemma}
\begin{proof} 
The sub-routine \CheckHiDegree($M,\eps$) selects a random subset $\bU \subset [n]$ of size $12/\eps$ by independently sampling from $[n]$, and random subset $\bE \subset [n]\times [n]$ of size $48n^{2/3}/\ea$ by repeatedly sampling from $[n]\times [n]$. For each index $i\in \bU$ and each pair of indices $(j,k)\in \bE$, the sub-routine checks whether the triangle $\{i,j,k\}$ is a violating triangle of $M$. If it is a violating triangle, the sub-routine outputs ``reject". Otherwise, it outputs ``accept". The first item follows directly from the procedure as any matrix in $\calP$ does not contain violating triangles. On the other hand, let $i\in[n]$ such that $d_T(i)\geq \ea n^{4/3}/16$ be called the high-degree indices. In order for the algorithm to output ``accept,'' it must avoid sampling a high-degree index, or if it does sample a high-degree index, must avoid the corresponding $\ea n^{4/3} / 16$ pairs of indices which form the violation. The probability this occurs is at most
\[ \left( 1 - \frac{\eps}{\sco}\right)^{12/\eps} + \left(1 - \frac{\ea}{16n^{2/3}} \right)^{48 n^{2/3}/\ea} \leq 1/6. \]
\end{proof}

\section{Proof from Section~\ref{sec:ultra} (Ultrametric and Tree Metric Testing Upper Bound)}

\begin{customlemma}{\ref{lem:large-EC-ODC}}
    Let $M\in\calC$ be $\eps$-far from $\calP^U$ and $S \subset [n]$ be consistent. If $|A(M, S)|\leq \eps n^2/8$, then either there are at least $\eps n/32$ inconsistent points in $[n]\setminus S$, or $|\SC(M, S)| + |\EC(M,S)| \geq \eps n^2/8$.
\end{customlemma}
\begin{proof}
    In Section~\ref{sec:ultra}, we have outlined the proof of Lemma~\ref{lem:large-EC-ODC}. In particular, we want to show the contra-positive of the statement is true: given $M \in \calC$ and a consistent $S \subset [n]$ which satisfies $|A(M, S)| \leq \eps n^2/8$, $|\SC(M,S)| + |\EC(M,S)| \leq \eps n^2/8$ and at most $\eps n / 32$ inconsistent points in $[n] \setminus S$, there exists an $n \times n$ matrix $\tilde{M} \in \calP^{U}$ which differs from $M$ on fewer than $\eps n^2$ entries. The matrix $\tM$ is constructed in Section~\ref{sec:ultra} and it is shown that $\|\tM-M\|_0<\eps n^2$.
    
    It remains to show that $\tilde{M}\in\calP^U$. We do this by considering an arbitrary triple $i, j, k \in [n]$, and showing that it does not form a violation in $\tilde{M}$. First, if at least one of three $i,j$ or $k$ had been inconsistent with $S$, then because we set distances from this point to be arbitrarily large, the triangle $i,j,k$ is no longer violated. A second easy case occurs when all $i,j,k \in S$. In this case, the consistency of $S$, and the fact $\tilde{M}_{|S\times S} = M_{|S \times S}$ implies there are no such violations.

    \begin{itemize}
        \item\label{en:two-in-S} Suppose that $i,j \in S$ and that $k \in [n] \setminus S$, and in addition, $k$ was consistent for $S$. In this case, all pairwise values among $\{i,j,k\}$ in $\tilde{M}$ are exactly the same as those in $M$. Since $k$ was consistent for $S$, the set $S \cup \{ k \}$ has no violations in $M$, and thus no violations in $\tilde{M}$.
        \item Suppose $i \in S$ and that $j, k \in [n] \setminus S$. We consider two sub-cases, according to whether or not $j$ and $k$ belong to the same part. 
        \begin{itemize}
            \item Suppose $j$ and $k$ belong to different parts. Then, Item~\ref{en:third-change} covers this case: If $(j,k) \notin \SC(M, S)$, then $M(j,k) = \max\{ M(i',j), M(i',k)\}$ for all $i' \in S$, and hence also for $i$. Since $\tilde{M}(j,k) = M(j,k)$, this is not a violation. If $(j,k) \in \SC(M,S)$, then $\tilde{M}(j,k) = \max\{ M(i', j), M(i', k)\}$ for some $i' \in \SEP(j,k)$ (which may be different from $i$). We assume without loss of generality that $M(i', k) > M(i', j)$. Note that $\{i, i', k\}$ and $\{i, i', j\}$ are not violating triples in $\tilde{M}$ as these are covered by Item~\ref{en:two-in-S}. Since $j,k$ are consistent with $S$, $\tilde{M}(i',j)=M(i',j),$ $\tilde{M}(i',k)=M(i',k)$ for all $i' \in S$. We will now show that the maximum of $\tilde{M}(j,k)$, $\tilde{M}(i,j)$ and $\tilde{M}(i,k)$ is not unique, i.e., either 
            \[ \text{(i) } \tilde{M}(j,k) = \max\left\{ \tilde{M}(i,j), \tilde{M}(i,k) \right\} \quad \text{or}\quad \text{(ii) }\tilde{M}(j,k) \leq \tilde{M}(i,j) = \tilde{M}(i,k). \]

            \begin{itemize}
        \item Suppose $M(i',i)< M(i',j) < M(i',k)$. First, notice that $\tilde{M}(j,k)=\max\{M(i',j),M(i',k)\}$ by the second bullet in Item~\ref{en:third-change}, which is equal to $M(i',k)=\tilde{M}(i',k)$ by assumption of this case. Because $\{i,i',k\}$ is not violating in $\tilde{M}$ and $\tilde{M}(i',k)=M(i',k)>M(i',i)=\tilde{M}(i',i),$ we must have $\tilde{M}(i',k)=\tilde{M}(i,k)$. Similarly, since the triple $\{i,i',j\}$ is not violating in $\tilde{M}$ and $\tilde{M}(i',j)=M(i',j)>M(i',i)=\tilde{M}(i',i)$, we must have $M(i',j)=\tilde{M}(i',j)=\tilde{M}(i,j)$. In summary, $\tilde{M}(j,k)=\tilde{M}(i',k)=\tilde{M}(i,k)$. By assumption, $M(i',k)>M(i',j)=\tilde{M}(i,j)$. Thus, (i) holds.
        \item Suppose $M(i',i)=M(i',j)<M(i',k)$. By the same first step, $\tilde{M}(j,k)=M(i',k)$. Since the triple $\{i,i',k\}$ is not violating in $M$ and $M(i',k)>M(i',i)$, we must have $M(i',k)=M(i,k)$. Moreover, since the triple $\{i,i',j\}$ is not violating in $M$ and $M(i',j)=M(i',i),$ we must have $M(i',j)\geq M(i,j)=\tilde{M}(i,j).$ By assumption, $M(i',k)>M(i',j)$, which is at least $M(i,j)=\tilde{M}(i,j).$ In summary, $\tilde{M}(j,k)=M(i,k)=\tilde{M}(i,k)$, and $\tilde{M}(j,k)=M(i',k)>\tilde{M}(i,j)$. Thus, (i) holds.
        \item Suppose $M(i',j)<M(i',i)<M(i',k)$. By the same first step, $\tilde{M}(j,k)=M(i',k)$. Since the triple $\{i,i',k\}$ is not violating in $M$ and $M(i',k)>M(i',i),$ we must have $M(i',k)=M(i,k)$. Similarly, since the triple $\{i,i',j\}$ is not violating in $M$ and $M(i',i)>M(i',j)$, we must have $M(i',i)=M(i,j)$. In summary, $\tilde{M}(j,k)=M(i',k)=M(i,k)=\tilde{M}(i,k)$, and by assumption, $M(i',k)>M(i',i)$ which is equal to $M(i,j)=\tilde{M}(i,j)$. Thus, (i) holds.
        
        \item Suppose $M(i',j)<M(i',i)=M(i',k)$. By the same first step, $\tilde{M}(j,k)=M(i',k)$. Since the triple $\{i,i',j\}$ is not violating in $M$ and $M(i',i)>M(i',j)$ by assumption, we must have $M(i',i)=M(i,j)$. Since the triple $\{i,i',k\}$ is not violating in $M$ and $M(i',k)=M(i',i)$, we must have $M(i',k)\geq M(i,k)$. In summary, we have $\tilde{M}(k,j)=M(i',k)=M(i,j)=\tilde{M}(i,j),$ and we have $M(i',k)\geq M(i,k)=\tilde{M}(i,k)$. Thus, (i) holds.
        
        \item Suppose $M(i',j)<M(i',k)<M(i',i).$ By the same first step, $\tilde{M}(j,k)=M(i',k)$. Since the triple $\{i,i',k\}$ is not violating in $M$ and $M(i',k)<M(i',i)$, we must have $M(i',i)=M(i,k)$. Similarly, since the triple $\{i,i',j\}$ is not violating in $M$ and $M(i',j)<M(i',i),$ we must have $M(i',i)=M(i,j)$. In summary, $\tilde{M}(k,j)=M(i',k)<M(i',i)$, which is equal to $M(i,k)=\tilde{M}(i,k)$ and also equal to $M(i,j)=\tilde{M}(i,j).$ Thus, (ii) holds.
    \end{itemize}
        Thus $\{i,j,k\}$ is not violating in $\tilde{M}.$
    \item If $j$ and $k$ belong to the same part $P$, we do a case analysis based on the category of $P$.
    \begin{itemize}
        \item Suppose $P$ is an Easy Part or an Active Part in $M$. Then $\tilde{M}(j,k)$ is set to be $\x$. Since $j,k$ are in the same part, $M(i',j)=M(i',k)$ for all $i'\in S$ so in particular, $M(i,j)=M(i,k)$. By definition, $\x$ is the minimum positive entry in $M$, so $M(i,j)=M(i,k)\geq \x=\tilde{M}(j,k)$. Since $j,k$ are consistent with $S$, by item~\ref{en:second-change}, $\tilde{M}(i,j)=M(i,j)$ and $\tilde{M}(i,k)=M(i,k)$. Thus, the triple $\{i,j,k\}$ is not violating in $\tilde{M}$.
        \item Suppose $P$ is a Versatile Part. Then $\tilde{M}(j,k)$ is set to be $M^\ast_P(j,k)\leq \min_{i'\in S}\{M(i',j)\}\leq M(i,j).$ Moreover, as $j,k$ are in the same part, $M(i',j)=M(i',k)$ for all $i'\in S$, so in particular $M(i,j)=M(i,k)$. By same reason as above, $\tilde{M}(i,j)=M(i,j)$ and $\tilde{M}(i,k)=M(i,k)$. In summary, $\tilde{M}(j,k)\leq \tilde{M}(i,j)=\tilde{M}(i,k)$, so the triple $\{i,j,k\}$ is not violating in $\tilde{M}$.
    \end{itemize}
        \end{itemize}
    \item Suppose all $i,j,k\in[n]\setminus S$. We consider three sub-cases, according to how many different parts they were in.
    \begin{itemize}
        \item If $i,j,k$ belonged to the same part $P$, we go into a case analysis on the category of $P$.
        \begin{itemize}
            \item Suppose $P$ is an Easy Part or an Active Part in $M$. Then $\tilde{M}(i,j),\tilde{M}(j,k),\tilde{M}(i,k)$ are all set to be $\x$. Thus, triple $\{j,k,l\}$ is not violating in $\tilde{M}.$ 
            \item Suppose $P$ is a Versatile Part. $\tilde{M}(i,j),\tilde{M}(j,k),\tilde{M}(i,k)$ are set to be $M^\ast_P(i,j),M^\ast_P(j,k),M^\ast_P(i,k)$ respectively. By definition of $M^\ast_P$, this square matrix encodes an ultrametric, so the triple $\{j,k,l\}$ is non-violating in $\tilde{M}$.
        \end{itemize}
        \item Suppose two of the indices $i,j$ belonged to the same part $P$ and $k$ belonged to a different part $P'$, then there exists a separator $v\in S$ such that $M(v,k)\neq M(v,i)=M(v,j)$. Moreover, as $i,j,k$ are consistent with $S$, we must have $\tilde{M}(v,k)=M(v,k)$, $\tilde{M}(v,i)=M(v,i)$, and $\tilde{M}(v,j)=M(v,j)$. Now consider the triple $\{v,k,i\}$: it is not violating in  $\tilde{M}$ because, by the argument above, any triple containing an index in $S$ is not violating in $\tilde{M}$. Thus, because of the inequality $\tilde{M}(v,k)=M(v,k)\neq M(v,i)=\tilde{M}(v,i)$, we must have $\tilde{M}(i,k)=\max\{\tilde{M}(v,k),\tilde{M}(v,i)\}.$ By the same argument, $\tilde{M}(j,k)=\max\{\tilde{M}(v,j),\tilde{M}(v,k)\}$. Thus, the chain of equality holds: 
        \[
        \tilde{M}(i,k)=\max\{\tilde{M}(v,k),\tilde{M}(v,i)\}=\max\{\tilde{M}(v,k),\tilde{M}(v,j)\}=\tilde{M}(j,k).
        \]
        Now, we go into a case analysis on the category of $P$.
        \begin{itemize}
            \item Suppose $P$ is an Easy Part or an Active Part. Then $\tilde{M}(i,j)$ is set to be $x$. Therefore, $\tilde{M}(i,k)=\tilde{M}(j,k)\geq \tilde{M}(v,k)$ by the chain of equality above, while $\tilde{M}(v,k)=M(v,k)\geq \x$ as $\x$ is the minimum positive entry in $M$. Thus, $\tilde{M}(i,k)=\tilde{M}(j,k)\geq \tilde{M}(i,j).$
            \item Suppose $P$ is a Versatile Part. Then $\tilde{M}(i,j)$ is set to be $M^\ast_P(i,j)$, which is at most $\min_{i'\in S,j\in P}(M(i',j))\leq M(v,j)=\tilde{M}(v,j)$. Therefore, $\tilde{M}(i,k)\leq \tilde{M}(v,j)\leq \tilde{M}(j,k)=\tilde{M}(i,k)$ by the chain of equality above.
        \end{itemize}
        All of the above show that $\{i,j,k\}$ are not-violating in $\tilde{M}$.
        \item Suppose $i,j,k$ belonged to three different parts $P_1,P_2,P_3$ respectively. Then there exists a separator $v\in \SEP(i,j)\subset S$ such that $M(v,i)\neq M(v,j)$. We will now assume, without loss of generality, that $M(v,i)<M(v,j)$ (by re-naming $i$ and $j$). By the above argument, $\{v,i,j\}$ contains one index in $S$ and so is not violating in $\tilde{M}$. Moreover, $\tilde{M}(v,i)=M(v,i)$, $\tilde{M}(v,j)=M(v,j)$, and $\tilde{M}(v,k)=M(v,k)$ as $i,j,k$ are consistent with $S$. Thus, we must have $\tilde{M}(i,j)=\max\{\tilde{M}(v,i),\tilde{M}(v,j)\}=M(v,j)$. Now, we go into a case analysis on how $M(v,k)$ compares with $M(v,i)$ and $M(v,j)$.
        \begin{itemize}
            \item Suppose $M(v,k)\neq M(v,i)$ and $M(v,k)\neq M(v,j)$. This implies $\tilde{M}(v,k)\neq\tilde{M}(v,i)$ and $\tilde{M}(v,k)\neq \tilde{M}(v,j)$ Then because triple $\{v,i,k\}$ contains an index in $S$, it is not violating in $\tilde{M}$ so we must have $\tilde{M}(i,k)=\max\{\tilde{M}(v,k), \tilde{M}(v,i)\}=\max\{M(v,k),M(v,i)\}$; similarly, triple $\{v,j,k\}$ is not violating in $\tilde{M}$ so we must have $\tilde{M}(k,j)=\max\{\tilde{M}(v,k), \tilde{M}(v,j)\}=\max\{M(v,k), M(v,j)\}$. If $M(v,k)>M(v,j),$ then $M(v,k)>M(v,i)$ as well by the assumption $M(v,j)>M(v,i)$, so $ \tilde{M}(i,k)=M(v,k)$ and $\tilde{M}(k,j)=M(v,k)$. Since $\tilde{M}(i,j)=M(v,j)<M(v,k)$, the triple $\{i,j,k\}$ is not violating in $\tilde{M}$. Now if $M(v,k)<M(v,j)$, then $\tilde{M}(k,j)=M(v,j)$. Since $\tilde{M}(i,k)=\max\{M(v,k),M(v,i)\}<M(v,j)$ and $\tilde{M}(i,j)=M(v,j),$ the triple $\{i,j,k\}$ is not violating in $\tilde{M}$.
            \item Suppose $M(v,k)=M(v,i)$. As $M(v,i)<M(v,j)$, $M(v,k)$ is less than $M(v,j)$ as well. These imply $\tilde{M}(v,k)=\tilde{M}(v,i)<\tilde{M}(v,j)$. Therefore, as the triple $\{v,k,j\}$ is not violating in $\tilde{M}$, we must have $\tilde{M}(k,j)=\tilde{M}(v,j)=M(v,j)$. Moreover, since the triple $\{v,i,k\}$ is not violating in $\tilde{M}$, we must have $\tilde{i,k}\leq \tilde{M}(v,k)=\tilde{M}(v,i)<\tilde{M}(v,j)=M(v,j).$ Therefore, the triple $\{i,j,k\} $ is not violating in $\tilde{M}$.
            \item Suppose $M(v,k)=M(v,j)$. As $M(v,i)<M(v,j)$, $M(v,k)$ is larger than $M(v,i)$ as well. These imply $\tilde{M}(v,i)<\tilde{M}(v,j)=\tilde{M}(v,k)$. Since the triple $\{v,i,k\}$ is not violating in $\tilde{M}$, we must have $\tilde{M}(i,k)=\tilde{M}(v,k)=\tilde{M}(v,j)=M(v,j)$. Since the triple $\{v,j,k\}$ is not violating in $\tilde{M}$, we must have $\tilde{M}(j,k)\leq \tilde{M}(v,j)=M(v,j)$. Together with $\tilde{M}(i,j)=M(v,j),$ the triple $\{i,j,k\}$ is not violating in $\tilde{M}$.
        \end{itemize}
    \end{itemize}
    \end{itemize}

    In summary, $\|M-\tM\|_0< \eps n^2$ and $\tM\in \calP^U$, which implies that the contra-positive of the lemma statement is true.
\end{proof}
\end{document}